\documentclass[a4paper, aps, prd,nofootinbib, notitlepage, 10pt]{revtex4-1}
\usepackage[OT1,T1]{fontenc}
\usepackage[latin1]{inputenc}
\usepackage{amsmath}
\usepackage{amsfonts}
\usepackage{bbm}
\usepackage{accents}
\usepackage{dsfont}
\usepackage{mathrsfs}
\usepackage[dvips]{graphicx}

\usepackage{color}
\usepackage{graphicx}
\newcommand{\qq}[1]{``#1''} 
\usepackage{hyperref}
\newenvironment{subalign}{\subequations\align}{\endalign\endsubequations}
\newcommand{\utilde}[1]{\undertilde{#1}}
\newcommand{\di}{\mathrm{d}} 
\newcommand{\ou}[3]{{#1}{}^{#2}{}_{#3}} 
\newcommand{\uo}[3]{{#1}{}_{#2}{}^{#3}} 
\newcommand{\I}{\mathrm{i}} 
\newcommand{\E}{\mathrm{e}} 
\newcommand{\ellp}{{\ell_{\mathrm{P}}}} 
\newcommand{\odens}{\widetilde{\eta}} 
\newcommand{\CC}{\mathrm{cc.}} 
\newcommand{\HC}{\mathrm{hc.}} 

\newcommand{\eref}[1]{(\ref{#1})}

\newcommand{\C}{\mathbb{C}}

\newcommand{\R}{\mathbb{R}}
\newcommand{\Z}{\mathbb{Z}}
\newcommand{\T}{\mathbb{T}}
\newcommand{\su}{{\mathfrak{su}}}

\renewcommand{\sl}{{\mathfrak{sl}}}

\renewcommand{\Im}{\mathfrak{Im}}
\newcommand{\bra}[1]{\langle{#1}|}
\newcommand{\ket}[1]{|{#1}\rangle}

\begin{document}
\title{Hamiltonian spinfoam gravity}
\author{Wolfgang M. Wieland}
\email{wolfgang.wieland@cpt.univ-mrs.fr}
\affiliation{Centre de Physique Théorique,
   Campus de Luminy, Case 907,
   13288 Marseille, France, EU}
\thanks{Unité Mixte de Recherche (UMR 7332) du CNRS et de l'Université 
d'Aix-Marseille et de l'Univ Sud Toulon Var. Unité affiliée à la
FRUMAM.}

\date{January 2013}
\begin{abstract}\noindent
This paper presents a Hamiltonian formulation of spinfoam-gravity, which leads to a straight-forward canonical quantisation. To begin with, we derive a continuum action adapted to the simplicial decomposition. The equations of motion admit a Hamiltonian formulation, allowing us to perform the constraint analysis. We do not find any secondary constraints, but only get restrictions on the Lagrange multipliers enforcing the reality conditions. This comes as a surprise. In the continuum theory, the reality conditions are preserved in time, only if the torsionless condition (a secondary constraint) holds true. Studying an additional conservation law for each spinfoam vertex, we discuss the issue of torsion and argue that spinfoam gravity may indeed miss an additional constraint. Next, we canonically quantise. Transition amplitudes match the EPRL (Engle--Pereira--Rovelli--Livine) model, the only difference being the additional torsional constraint affecting the vertex amplitude. 
\end{abstract}
\maketitle
\section{The problem}\noindent
Loop quantum gravity \cite{ashtekar, rovelli, thiemann, status, alexreview} comes in two versions. The historically first of which provides a canonical quantisation of general relativity, and seeks to solve the Wheeler--De Witt equation \cite{ashtekar, thiemann, inside}. The second, we call it spinfoam gravity \cite{ alexreview, reisenberger}, proposes a covariant path integral. Both approaches share \cite{physbound, covol} their kinematical structure---the Hilbert space with observables representing, area, angles, volume and parallel transport \cite{Rovelliarea, sethangles, ashvolume}.
Unfortunately, we know very little about how far this relation extends beyond kinematics, eventually revealing a solid framework also for the dynamics of the theory. Indications supporting this idea have only come from the symmetry reduced framework of loop quantum cosmology \cite{lqcspinfoam2, lqcspinfoam3, lqcspinfoam4}.

The problem of how the canonical theory and spinfoam gravity can fit together is not only of mathematical significance. First of all, it is a consistency check for the theory. If loop gravity were a fully developed theory, it should come in different formulations, providing specific advantages and simplifications, although being mutually equivalent. On top of that, a framework large enough to contain both Lagrangian and Hamiltonian dynamics should allow us to answer some of the most pressing questions in the field: What is the classical theory underlying loop gravity, and does it reproduce general relativity? Does loop gravity contain torsion, and are there any secondary constraints missing? Is there a local notion of energy, and what can we use it for?

This paper achieves such a unifying framework in the simplified setting of a fixed discretisation. We will, in fact, develop a Hamiltonian formulation of spinfoam gravity, adapted to a simplicial decomposition of space-time. This discretisation consists of triangles, tetrahedra and 4-simplices glued together. Calculations will heavily use the twistorial framework of loop quantum gravity, recently developed by Dupuis, Freidel, Livine, Speziale, Tambornino and myself \cite{twist, twistintegrals, komplexspinors, twistconslor, holsimpl, spezialetwist1}. These spinorial variables do not change the physical content of loop gravity, but offer a new perspective on how to look at the covariant aspects of the theory. 

The article has three parts. The first of which, section \ref{chapterII}, is a technical review, necessary to develop the main results. We speak about selfdual 2-forms, complex variables, reality conditions, and discuss the spinorial parametrisation of loop gravity. Section \ref{chapterIII} contains the classical part, we introduce a Hamiltonian for spinfoam gravity, study the equations of motion, and perform the constraint analysis. Working with complex variables we have to deal with reality conditions, which guarantee that the metric is real. The Hamiltonian time evolution must preserve these constraint equations, a condition that may need additional secondary constraints to be fulfilled. We will, however, not find any secondary constraints, but only get restrictions on the Lagrange multipliers. This comes as a surprise. For, as we know from the Hamiltonian analysis of the continuum theory, the torsionless condition turns into a secondary constraint needed to preserve the reality conditions. If we do not get any secondary constraints from our canonical analysis, where is the torsionless condition hiding? At the end of section \ref{chapterIII} we will discuss this issue, and argue spinfoam gravity may indeed miss an additional torsional constraint. 
Section \ref{chapterIV} concerns quantum theory. The Hamiltonian previously introduced turns into an operator generating the local time evolution around a spinfoam face. Solving the constraint equations, we find transition amplitudes that share key features of the EPRL (Engle--Pereira--Rovelli--Livine) model \cite{zakolec, flppdspinfoam, LQGvertexfinite}. When looking at a spinfoam face both amplitudes agree, the only difference concerns the way these amplitudes glue to form a spinfoam.

\section{Selfdual 2-forms, spinors and reality conditions for loop gravity}\label{chapterII}
\noindent To make this paper logically self-contained this section briefly recapitulates results from previous publications \cite{komplex1, komplexspinors, twistintegrals, realcond}. Section \ref{sectionIIA} discusses complex Ashtekar variables and the discrete holonomy-flux algebra they induce. In section \ref{sectionIIB} we review the twistorial parametrisation of the reduced phase-space of holonomy-flux variables, while section \ref{sectionIIC} gives the reality conditions.
\subsection{Complex Ashtekar variables}\label{sectionIIA}
\noindent The spinfoam approach seeks to define transition amplitudes for loop quantum gravity. It starts from the following topological action:
\begin{equation}
S[\Sigma,A]=\frac{\I\hbar}{\ellp^2}\frac{\beta+\I}{\beta}\int_{\mathcal{M}}\Sigma_{AB}\wedge F^{AB}(A)+\CC,\label{bfactn}
\end{equation}
which is the \qq{$BF$-action} \cite{perez} expressed in selfdual variables.
Here $\ellp^2=8\pi\hbar G/c^3$ is the Planck area, $\beta$ is the Barbero--Immirzi parameter, $F=\di A+\frac{1}{2}[A,A]$ is the curvature of the selfdual connection, $\ou{\Sigma}{A}{B}$ denotes an $\mathfrak{sl}(2,\C)$ valued 2-form, and the antisymmetric $\epsilon$-tensor\footnote{The $\epsilon$-tensor lowers indices as $v_A=v^B\epsilon_{BA}\in\C^2{}^\ast$, while its inverse raises them by $v^A=\epsilon^{AB}v_B\in\C^2{}$; the inverse is implicitly defined by putting $\epsilon_{AC}\epsilon^{BC}=\uo{\epsilon}{A}{B}=\delta^B_A$.} moves the spinor indices $A,B,C,\dots\in\{0,1\}$. These indices transform under the fundamental $(\tfrac{1}{2},0)$ representation of $SL(2,\C)$. We use \qq{cc.} to denote the complex conjugate of everything preceding (including the pre-factor $\I(\beta+\I)\dots$), and so the $(0,\tfrac{1}{2})$ representation also appears. Indices transforming under this complex conjugate representation carry an overbar, we write $\bar A,\bar B,\bar C,\dots$. Working with a closed manifold, we do not have to worry about boundary terms that are otherwise needed \cite{surholst}.

This action shares the symplectic structure of general relativity, but the dynamics is trivial. Indeed, performing a 3+1 split $\mathcal{M}=\mathbb{R}\times\mathcal{S}\ni(t,p)$, we find the symplectic structure of complex Ashtekar variables:
\begin{equation}
\big\{\uo{\Pi}{i}{a}(p),\ou{A}{j}{b}(q)\big\}=\delta^j_i\delta^a_b\tilde\delta(p,q)=
\big\{\uo{\bar\Pi}{i}{a}(p),\ou{\bar A}{j}{b}(q)\big\},\label{poissbrack}
\end{equation}
here indices $i,j,k$ running from 1 to 3 refer to the standard basis\footnote{Given any $\phi\in\mathfrak{sl}(2,\C)$ we write $\phi=\phi^i\tau_i$, where $\sigma_i=2\I\tau_i$ are the Pauli matrices.} in $\mathfrak{sl}(2,\C)$, $\tilde\delta$ is the Dirac-delta density on the spatial hypersurface $\mathcal{S}_t=\{t\}\times\mathcal{S}$, and $a,b,c,\dots$ are abstract indices on the spatial slice. The Ashtekar connection $\ou{A}{i}{a}=\ou{\Gamma}{i}{a}+\I\ou{K}{i}{a}$ is the pullback of the selfdual connection onto the spatial hypersurfaces; in general relativity its real and imaginary parts ($\Gamma$ and $K$ respectively) correspond to the intrinsic $\mathfrak{su}(2)$ connection and the extrinsic curvature of the spatial hypersurface. The momentum conjugate is linearly related to $\Sigma$, we have:
\begin{equation}
\uo{\Pi}{i}{a}=-\frac{\hbar}{\ellp^2}\frac{\beta+\I}{2\I\beta}\uo{\Sigma}{i}{a}=-\frac{\hbar}{\ellp^2}\frac{\beta+\I}{4\I\beta}\odens^{abc}\Sigma_{ibc},\label{pidef}
\end{equation}
with $\odens^{abc}$ the Levi-Civita density on the spatial hypersurface.

The continuum Poisson brackets behave too singularly to perform a background-independent quantisation. Therefore, we introduce a reduced phase-space of smeared variables. We can define the smeared variables most elegantly when we consider a cellular decomposition of the $t=\mathrm{const.}$ hypersurface. In this paper, we restrict to triangulations, and thus divide the spatial manifold into tetrahedra glued among bounding faces. Generalisations to arbitrary cellular decompositions exist and have been studied elsewhere, e.g. in \cite{EEcomm}.

Within the spatial manifold, the faces, the oriented triangles $\tau_1$, $\tau_2,\dots, \tau_L$, are the duals of oriented links $\gamma_1,\gamma_2,\dots \gamma_L$. 
To smear the connection, we take a link and study the $SL(2,\C)$ parallel transport between adjacent tetrahedra. We are thus led to the holonomy:
\begin{equation}
h[\tau]=\mathrm{Pexp}\big(-\int_\gamma A\big)\in SL(2,\C).
\end{equation} 
The momentum variable $\Pi$ defines a 2-form; parallel transported into the frame of a tetrahedron, we can naturally smear it over an adjacent triangle $\tau$ obtaining the gravitational flux:
\begin{equation}
\Pi[\tau]=\int_{p\in \tau}h_{\delta(p\rightarrow\gamma(0))}^{-1}\Pi_p h_{\delta(p\rightarrow\gamma(0))}\in\mathfrak{sl}(2,\C),\label{gravfluxdef}
\end{equation}
where
$h_{\delta(p\rightarrow\gamma(0))}$ is an $SL(2,\C)$ holonomy connecting $p\in \tau$ with the source point $\gamma(0)$. The underlying path $\delta(p\rightarrow\gamma(0))$ consist of two parts, the first of which lies inside $\tau$ and goes from $p\in \tau$ towards the intersection point $\tau\cap\gamma$, whereas the second part goes from the intersection point along $\gamma$ towards the source $\gamma(0)$.

The continuum Poisson brackets \eref{poissbrack} induce commutation relations among holonomies and fluxes; variables belonging to different triangles commute, while for a single link we get the commutation relations of $T^\ast SL(2,\C)$:
\begin{equation}
\big\{\Pi_i,\Pi_j\big\}=-\uo{\epsilon}{ij}{k}\Pi_k,\quad
\big\{\Pi_i,h\big\}=-h\tau_i,\quad
\big\{\ou{h}{A}{B},\ou{h}{C}{D}\big\}=0.\label{holfluxal}
\end{equation}
There are also the Poisson brackets of the anti-selfdual variables, which are nothing but the complex conjugate of the former variables.
Moreover, just as in \eref{poissbrack}, the two sectors of opposite chirality commute. Since $\tau$ carries an orientation let us also mention the quantities:
\begin{equation}
h[\tau^{-1}]=h[\tau]^{-1},\quad \utilde{\Pi}[\tau]:=\Pi[\tau^{-1}]=-h[\tau]\Pi[\tau]h[\tau]^{-1}.
\end{equation}

Before we go on let us make one more observation. The definition of the flux  \eref{gravfluxdef} depends on the underlying family of paths $\delta(p\rightarrow \gamma(0))$ chosen. It is therefore quite remarkable \cite{EEcomm} that this dependence drops out of the Poisson algebra \eref{holfluxal}, and leads us to the phase space $T^\ast SL(2,\C)$.  
\subsection{Spinors for loop gravity}\label{sectionIIB}
\noindent The phase space of loop gravity on a graph, $T^\ast SL(2,\C)^L$,  allows for a description in terms of spinors. This framework will become important for the rest of the paper; it is useful for us since it embeds the non-linear phase space $T^\ast SL(2,\C)^L$ into a vector space with canonical Darboux coordinates.

The flux defines an $\mathfrak{sl}(2,\C)$ element, which is traceless, and we can thus always find a pair of diagonalising spinors (the proof can be found in \cite{penroserindler1}, which together with \cite{komplexspinors}, can serve as the main reference of this section):
\begin{equation}
\Pi[\tau]_{AB}=-\frac{1}{2}\omega_{(A}\pi_{B)}=-\frac{1}{4}\big(\omega_A\pi_B+\omega_B\pi_A\big).\label{pidef2}
\end{equation}
With the flux evaluated in the frame of the source $\gamma(0)$, also $\pi$ and $\omega$ belong to the initial point. If we parallel transport them to the target point we get another pair of spinors:
\begin{equation}
\utilde{\pi}^A=\ou{h[\tau]}{A}{B}\omega^B,\quad\utilde{\omega}^A=\ou{h[\tau]}{A}{B}\pi^B.\label{holdef}
\end{equation}
Since the spinors $\pi$ and $\omega$ often come as a pair, it is useful to introduce the twistor:
\begin{equation}
Z:=(\bar{\pi}_{\bar A},\omega^A)=(\bar\C^2)^\ast\oplus\C^2=:\T
\end{equation}
If $\pi$ and $\omega$ be linearly independent, that is 
\begin{equation}
\pi_A\omega^A\neq 0,
\end{equation}
they form a basis in $\C^2$. We can safely agree on this constraint because it holds true unless the triangle represents a null surface, and we are working with spatial hypersurfaces anyhow. In this case, equation \eref{holdef} gives the holonomy in a certain basis, which uniquely fixes this $SL(2,\C)$ element. We can now reverse the construction, start with a pair $(\utilde{Z},Z)$ of twistors and attach them to the source and target point respectively.
Inverting equations \eref{holdef} and \eref{pidef2} we then recover both holonomy and flux in terms of spinors. The holonomy explicitly reads
\begin{equation}
\ou{h[\tau]}{A}{B}=\frac{\utilde{\omega}^A\pi_B-\utilde{\pi}^A\omega_B}{\sqrt{\pi\omega}\sqrt{\utilde{\pi}\utilde{\omega}}},\quad \text{with: $\pi\omega:=\pi_A\omega^A$}\label{holdef2}.
\end{equation}
For the holonomy to have unit determinant it must preserve $\epsilon_{AB}$. Within our spinorial framework this immediately turns into the \emph{area-matching constraint}
\begin{equation}
C[\utilde{Z},Z]=\utilde{\pi}_A\utilde{\omega}^A-\pi_A\omega^A=0.\label{armtch}
\end{equation}
The space of spinors on a link can be equipped with a locally $SL(2,\C)$ invariant symplectic structure. We set
\begin{subalign}
\big\{\pi^A,\omega^B\big\}&=\epsilon^{AB}=-\big\{\utilde{\pi}^A,\utilde{\omega}^B\big\}\label{commtwist}\\
\big\{\bar\pi^{\bar A} ,\bar\omega^{\bar B}\big\}&=\bar\epsilon^{\bar A\bar B}=-\big\{\utilde{\bar\pi}^{\bar A},\utilde{\bar\omega}^{\bar B}\big\},
\end{subalign}
and can prove \cite{komplexspinors} that on the constraint hypersurface $C=0$, these Poisson brackets induce the commutation relations of $T^\ast SL(2,\C)$ for the flux \eref{pidef2} and the holonomy \eref{holdef2}. The parametrisations \eref{holdef2} and \eref{pidef2} are not unique. There are in fact two symmetries; one of which is discrete, the other is continuous. The discrete symmetry simultaneously exchanges $\pi$ with $\omega$, and $\utilde{\pi}$ with $\utilde{\omega}$, while the Hamiltonian vector field of $C$ generates the conformal symmetry
\begin{equation}
(\omega,\pi;\utilde{\omega},\utilde{\pi})\mapsto (z\omega,z^{-1}\pi;z\utilde{\omega},z^{-1}\utilde{\pi}),\quad z\in\C-\{0\}\label{confsym},
\end{equation}
leaving both \eref{holdef2} and \eref{pidef2} invariant.
We perform the symplectic quotient with respect to these symmetries and eventually get $T^\ast SL(2,\C)$ removed from all its null configurations $(\Pi,h)\in T^\ast SL(2,\C):\Pi_{AB}\Pi^{AB}=0$. The proof is again in \cite{komplexspinors}. 

\subsection{Reality conditions}\label{sectionIIC}
\noindent The action \eref{bfactn} defines a topological theory. We recover general relativity in terms of first-order variables only if we impose constraints on $\Sigma_{AB}$. We want $\Sigma_{AB}$ to be geometrical, that is to represent an infinitesimal area element. In the continuum theory this means $\Sigma_{AB}$ should be the selfdual part of the Plebanski 2-form $\Sigma_{\alpha\beta}=e_\alpha\wedge e_\beta$ (where $e^\alpha$ is the tetrad and $\alpha=0,\dots,3$ are internal Minkowski indices). 

The Hamiltonian flow preserves these \emph{simplicity constraints} \cite{selfdualtwo, flppdspinfoam, LQGvertexfinite} only if the space-time connection is torsionless (which can be shown in many ways, e.g.  \cite{komplex1, realcond, Alexandrovanalysis}). In the Hamiltonian framework, equations of motion either are constraints (possibly both on Lagrangian multipliers and the phase-space) or evolution equations. Conversely, the torsionless condition of the space-time connection splits into three distinct parts \cite{komplex1, Geillertors2}. Using time gauge (thus aligning the internal normal to the hypersurface normal, that is setting $e^0=N\di t$), we find the Lagrange multiplier\footnote{In the following $\partial_t$ denotes the time-flow vectorfield.} $\Lambda^i=\Im{(A^i(\partial_t))}$ is determined by the lapse ($N$) and the shift ($N^a$) to have the value $\Lambda^i=N^a\ou{K}{i}{a}+e^{ia}\partial_a N$. The second part of the torsionless condition represents an evolution equation for the spatial triad, whereas the last and most important part gives a second-class constraint among the phase-space variables:
\begin{equation}
\ou{A}{i}{a}+\CC=2\ou{\Gamma}{i}{a}(e),\label{trslss}
\end{equation}
where $\Gamma(e)$ is the spatial Levi-Civita connection functionally depending on the triad.

In the discrete theory the situation is different. We do not have a continuous tetrad, and we cannot define the continuous simplicity constraints $\Sigma_{\alpha\beta}-e_\alpha\wedge e_\beta=0$ directly. Instead we have smeared variables on a triangulation of a $t=\mathrm{const.}$ spatial hypersurface. However, the physical meaning of the simplicity constraints is clear, they guarantee that the 2-form $\Sigma_{\alpha\beta}$ is geometric, hence defines a plane in internal space. For our smeared variables we can demand something similar: $\Sigma_{\alpha\beta}[\tau]$, that is the 2-form $\Sigma_{\alpha\beta}$ smeared over the triangle $\tau$ in the frame of the tetrahedron at $\gamma(0)$, should define a spatial plane in internal Minkowski space. This is true if there is a time-like vector $n^\alpha$ such that:
\begin{equation}
\Sigma_{\alpha\beta}[\tau]n^\beta\equiv-\Sigma_{AB}[\tau]\bar\epsilon_{\bar A\bar B}n^{B\bar B}-\CC=0.\label{simpl1}
\end{equation}
These are the linear simplicity constraints \cite{flppdspinfoam}, which are reality conditions on the momentum variables \cite{komplex1}.
Geometrically, the vector $n^\alpha$ represents the normal to the tetrahedron the triangle $\tau$ is seen from. This normal should thus be the same for all four triangles meeting at a tetrahedron. A detailed discussion of the geometric origin of the simplicity constraints can be found in the appendix of reference \cite{flppdspinfoam}.

In terms of spinors, equation \eref{simpl1} turns into two independent constraints \cite{komplex1}. One is real, the other is complex, and there are thus three real constraints to solve:
\begin{subequations}
\begin{align}
D[\omega,\pi]&=\frac{\I}{\beta+\I}\pi_A\omega^A+\CC=0,\\
F_n[\omega,\pi]&=n^{A\bar A}\pi_A\bar\omega_{\bar A}=0.
\end{align}\label{simplspin}
\end{subequations}
The constraint $D=0$, is locally Lorentz invariant and guarantees the area of $\tau$ is real; $F_n=0$, on the other hand, is preserved only under spatial rotations, and tells us the null-vector $m^\alpha\equiv\omega^A\bar\pi^{\bar A}$ in complexified Minkowski space lies orthogonal to $n^\alpha$.
If we were to work in the time-gauge, we would align the normal $n^\alpha$ with $n^\alpha_o=\delta^\alpha_0$ and the matrix $n^{A\bar A}$ would turn into
\begin{equation}
n^{A\bar A}=\frac{\I}{\sqrt{2}}\delta^{A\bar A}\equiv n_o^{A\bar A},\label{tgauge}
\end{equation}
where $\delta^{A\bar A}$ is the identity matrix. In this gauge the real and imaginary parts of the momentum
\begin{equation}
\Pi_i=\frac{1}{2}\big(L_i+\I K_i\big).
\end{equation}
generate rotations and boosts, and the reality conditions \eref{simplspin} turn into:
\begin{equation}
\frac{1}{\beta+\I}\Pi_i+\CC=\frac{1}{\beta^2+1}\big(K_i+\beta L_i\big)=0.\label{linearsimpl}
\end{equation} 

Whether or not spinfoam gravity misses the secondary constraints, and forgets about equation \eref{trslss}, raises some of the most pressing and strongly debated \cite{PhysRevD.82.064026, Dittrich:2012rj, Geillertors, sergeiarea, oritigft} questions in our field. The spinorial framework of loop quantum gravity will allow us to study this issue. In the next section we are going to prove that spinfoam gravity correctly implements the simplicity constraints without missing any secondary constraints. Although this sounds very promising for the model \cite{flppdspinfoam, LQGvertexfinite}, we will argue, at the end of the next section, that spinfoam gravity may still miss an important constraint related to torsion.

\section{Hamiltonian dynamics for spinfoam gravity}\label{chapterIII}\noindent
This section introduces a continuous formulation of the dynamics on a fixed 2-complex. We will check if the equations of motion preserve the reality conditions  (in whatever form, i.e. \eref{linearsimpl} or \eref{simplspin}, preferred), and study the equations of motion. They will, in fact, immediately prove the curvature smeared over a spinfoam face does not vanish, hence the model carries curvature. To achieve these claims, and properly answer the issue of the secondary constraints we need an action, or even better to find a suitable Hamiltonian framework. And this is what we are going to do first.
\subsection{The discrete action on a spinfoam wedge}
\begin{figure}
     \centering
     \includegraphics[width= 0.60\textwidth]{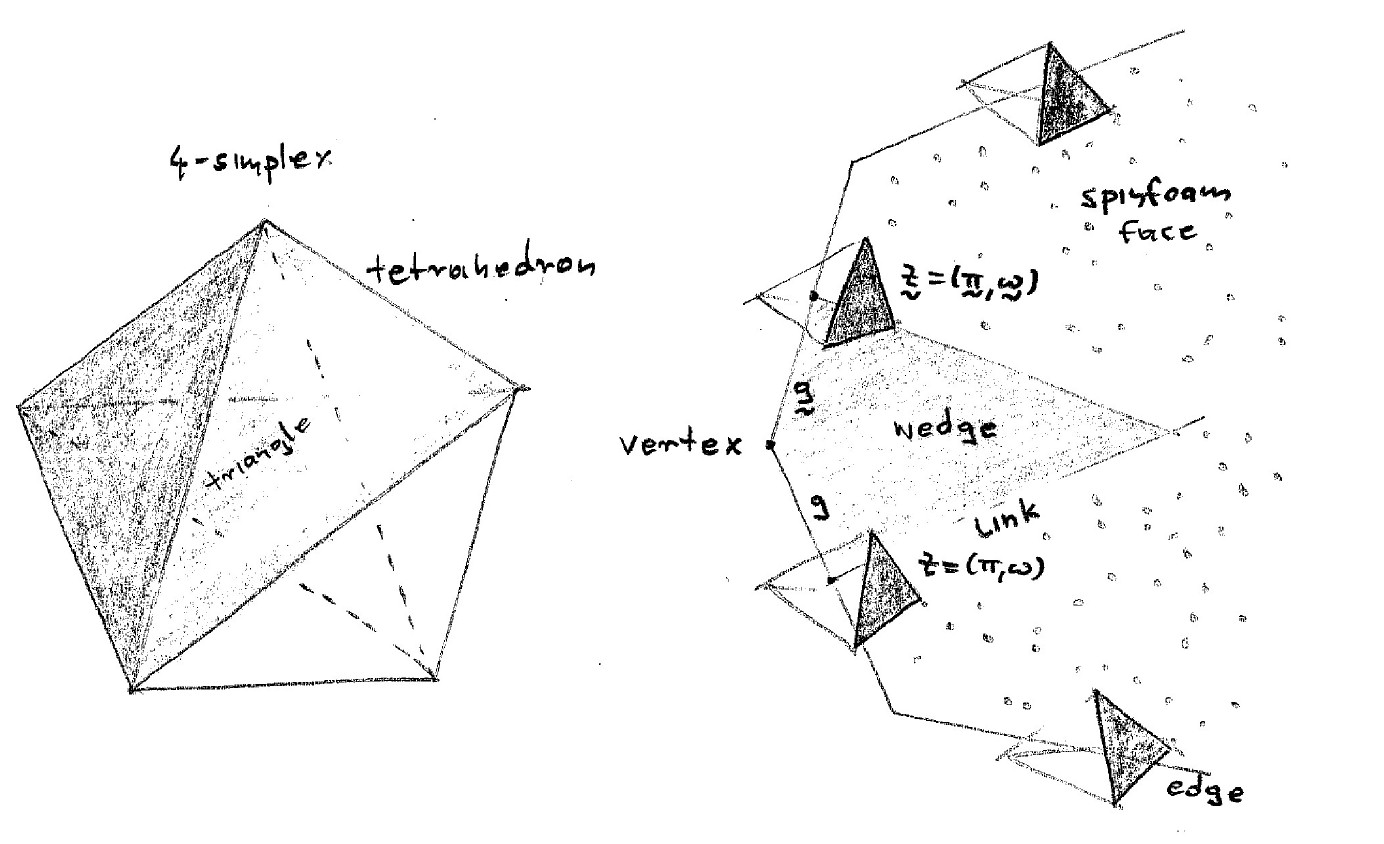}
     \caption{A 4-simplex consists of five tetrahedra glued among their triangles. Its dual we call a \emph{vertex}. Each triangle belongs to many 4-simplices (vertices), but a tetrahedron can only be in two of them. The surface dual to a triangle is a \emph{spinfoam face}, it touches all adjacent 4-simplices. An edge, a tetrahedron's dual, connects two vertices. The part of a spinfoam face lying inside a given 4-simplex, we call a \emph{spinfoam wedge}, the boundary of which has two parts. The first one consists of edges, enters the bulk, and passes through the vertex. The second part belongs to the boundary of the 4-simplex, we call it a link.}
     \label{simplx}
\end{figure}

\noindent 
We start with the topological action \eref{bfactn} discretised over a simplicial decomposition of the four-dimensional space-time manifold $\mathcal{M}$, and use our spinors to parametrise the action. 
 First steps towards this task have already been reported in \cite{twistintegrals}. The elementary building blocks are 4-simplices glued among the bounding tetrahedra. All tetrahedra consist of four triangles, each of which is dual to a spinfoam face. The part of a spinfoam face belonging to a 4-simplex we call a wedge $w$, for the corresponding dual triangle (in the frame of a tetrahedron) we write $\tau_w$. Figure \ref{simplx} should further clarify the geometry. 
Here we are committing ourselves to simplicial discretisations, which we do for technical convenience; generalisation to cellular decompositions should be found along the lines of \cite{Lewandowskigeneral}.

Let $F^{AB}[w]$ be the curvature tensor integrated over the wedge, and $\Sigma_{AB}[\tau_w]$ be the 2-form $\Sigma_{AB}$ smeared over the dual triangle (in the frame of a tetrahedron), while $w^{-1}$ and $\tau^{-1}$ denote the oppositely oriented surfaces. The remaining ambiguity concerns the relative orientation  $\epsilon(\tau_w,w)$ between the two surfaces, which we take to be one\footnote{If the tangent vectors $(x,y)$ are positively oriented in $\tau_w$, and the pair $(t,z)$ is positively oriented in $w$, the relative orientation $\epsilon(\tau_w,w)$ is the orientation of the quadruple $(x,y,t,z)$.}. 

We discretise the topological action over each 4-simplex and find a sum over wedges:
\begin{equation}
\begin{split}
S^{\mathrm{top}}[\Sigma,A]&=\frac{\I\hbar}{\ellp^2}\frac{\beta+\I}{\beta}\int_M\Sigma_{AB}\wedge F^{AB}(A)+\CC\\
&\approx \frac{\I\hbar}{\ellp^2}\frac{\beta+\I}{2\beta}\sum_{w:\text{wedges}}\big(\Sigma_{AB}[\tau_w]F^{AB}[w]+\Sigma_{AB}[\tau_w^{-1}]F^{AB}[w^{-1}]\big)+\CC\equiv\sum_{w:\text{wedges}}S_w^{\mathrm{top}}.
\end{split}\label{discrt}
\end{equation}
For small curvature we can replace $F^{AB}[w]$ by the holonomy around the loop $\partial w$ bounding the wedge:
\begin{equation}
h^{AB}[\partial w]=\mathrm{Pexp}\big(-\int_{\partial w}A\big)^{AB}\approx -\epsilon^{AB}+\int_w F^{AB}=:-\epsilon^{AB}+F^{AB}[w].
\end{equation}
Within this approximation we will now rewrite everything in terms of our spinorial variables. For the flux the situation is simple. The triangle belongs to a tetrahedron in the boundary of the 4-simplex. Boundary variables are part of the original phase-space $T^\ast SL(2,\C)$ of complex Ashtekar variables, and we can thus use our spinorial parametrisation \eref{pidef2} to write:
\begin{equation}
\Sigma_{AB}[\tau_w]=\frac{\ellp^2}{\hbar}\frac{\I\beta}{\beta+\I}\omega_{(A}\pi_{B)}.
\end{equation}
For the holonomy around the wedge we have to be more careful. The boundary of the wedge consists of two parts, one of which enters the bulk. The first part, connecting the two adjacent tetrahedra $\mathcal{T}$ and $\utilde{\mathcal{T}}$, lies in the boundary of the 4-simplex. The corresponding holonomy $h[\tau_w]$ is again contained in the phase-space of $T^\ast SL(2,\C)$, and we can thus take the spinorial parametrisation \eref{holdef2} to write the group element. The second part enters the bulk, and we need additional $SL(2,\C)$ elements $g$ and $\utilde{g}$ that give the parallel transport from the center of the 4-simplex towards $\mathcal{T}$ and $\utilde{\mathcal{T}}$ respectively. These additional holonomies are not part of our phase-space of complex Ashtekar variables, instead they are Lagrange multipliers, which should become clear once we discuss the Gauß law.
Gluing the two holonomies together we find the parallel transport around the boundary of the wedge starting at the tetrahedron $\mathcal{T}$:
\begin{equation}
h^{AB}[\partial w]=\big(g\utilde{g}^{-1}\big){}^A{}_Ch^{CB}[\tau_w]=\big(g\utilde{g}^{-1}\big){}^A{}_C\frac{\utilde{\omega}^C\pi^B-\utilde{\pi}^C\omega^B}{\sqrt{\utilde{\pi}\utilde{\omega}}\sqrt{\pi\omega}}
\end{equation}
We thus find the contribution to the discretised action from a single wedge to be:
\begin{equation}
S_w^{\mathrm{top}}=-\frac{1}{2}M_w\big(g\utilde{g}^{-1}\big)^{AB}\big(\omega_A\utilde{\pi}_B+\pi_A\utilde{\omega}_B\big)+\CC,\label{wactn1}
\end{equation}
where we have introduced the quantity: 
\begin{equation}
M_w=\frac{1}{2}\Big(\frac{\sqrt{\pi\omega}}{\sqrt{\utilde{\pi}\utilde{\omega}}}+\frac{\sqrt{\utilde{\pi}\utilde{\omega}}}{\sqrt{\pi\omega}}\Big).
\end{equation}
This normalisation equates to one once we go to the solution space of the area-matching constraint \eref{armtch}, where the action \eref{wactn1} turns into a simple bilinear of the spinors. The reality conditions \eref{simplspin} also decomposes into a bilinear of the spinors, and this is the reason why the spinorial parametrisation of loop quantum gravity will be so useful.
\subsection{The continuum action on a wedge}
\noindent The action \eref{wactn1} just introduced admits a straight forward continuum limit.  To show this, we split the wedge $w$ into smaller wedges $w_1,\dots, w_N$, introduce $N-1$ additional spinorial variables $(\omega^{(i)},\pi^{(i)})$ together with group elements $g^{(i)}\in SL(2,\C)$ that represent the parallel transport from the vertex to the $i$-th discretisation step at the boundary of the spinfoam face. For the $i$-th wedge the action \eref{wactn1} becomes:
\begin{equation}
S_{w_i}^{\mathrm{top}}=-\frac{1}{2}M_{w_i}\big(g^{(i)}(g^{(i+1)})^{-1}\big)^{AB}\big(\omega_A^{(i)}{\pi}^{(i+1)}_B+\pi_A^{(i)}\omega^{(i+1)}_B\big)+\CC,\label{wactn2}
\end{equation}
and we also have the boundary conditions  $(\omega^{(1)},\pi^{(1)},g^{(1)})=(\omega,\pi,g)$; $(\omega^{N+1},\pi^{(N+1)},g^{(N+1)})=(\utilde{\omega},\utilde{\pi},\utilde{g})$.
We now take the continuum limit $N\rightarrow\infty$. Put $\varepsilon=N^{-1}$, set all variables $f(\varepsilon(i-1)):=f^{(i)}$, and choose the quantity $t=\varepsilon(i-1)$ as our natural continuous time variable. 
To obtain the continuum limit we perform an expansion in $\varepsilon$ implicitly assuming all quantities $(\pi(t),\omega(t),g(t))$ are differentiable in the parameter time $t$.

Let us first look at the normalisation $M_{w_i}$. Putting $E(t)=E^{(i)}=\epsilon^{AB}\pi_A^{(i)}\omega_B^{(i)}$ we have
\begin{equation}
M_{w_i}=\frac{1}{2}\bigg(\frac{\sqrt{E(t)}}{\sqrt{E(t+\varepsilon)}}+\frac{\sqrt{E(t+\varepsilon)}}{\sqrt{E(t)}}\bigg)=1+\mathcal{O}(\varepsilon^2).
\end{equation}
We see the first non-vanishing order is quadratic. For the holonomies, on the other hand, the expansion contains a linear term:
\begin{equation}
g^{(i+1)}(g^{(i)})^{-1}=\mathrm{Pexp}\Big(-\int_{\varepsilon(i-1)}^{\varepsilon i}\di t\,A_{e(t)}(\partial_t)\Big)=\mathds{1}-\varepsilon A_{e(t)}(\gamma_t)+\mathcal{O}(\varepsilon^2).\label{holprod}
\end{equation}
Here $e$ is the path (the \qq{edge}) bounding the spinfoam face, and $t$ is the associated coordinate. 
Next, we need to study the product of the spinors. We find
\begin{equation}
\omega_A^{(i)}{\pi}^{(i+1)}_B=\omega_A\big(\varepsilon(i-1)\big)\pi_B(\varepsilon i)=\omega_A(t)\pi_B(t+\varepsilon)=\omega_A(t)\pi_B(t)+\varepsilon \omega_A(t)\dot{\pi}_B(t)+\mathcal{O}(\varepsilon^2).\label{spinprod}
\end{equation}
Combining \eref{spinprod} and \eref{holprod} we get the expansion of the bilinear appearing in the action:
\begin{equation}
\begin{split}
-\big(g^{(i)}(g^{(i+1)})^{-1}\big)^{AB}\omega_A^{(i)}{\pi}^{(i+1)}_B&=
\epsilon^{AB}\omega_A(t)\pi_B(t)+\varepsilon\epsilon^{AB}\omega_A(t)\dot{\pi}_B(t)-\varepsilon A^{AB}_{e(t)}(\dot e)\omega_B(t)\pi_A(t)+\mathcal{O}(\varepsilon^2)=\\
&=\omega_A(t)\pi^A(t)+\varepsilon\omega_A(t)D_{\partial_t}\pi^A(t)+\mathcal{O}(\varepsilon^2).
\end{split}
\end{equation}
The same is true for the second part of \eref{wactn2} with $\omega$ and $\pi$ exchanged.
Moreover, $D_{\partial_t}\pi^B=\dot\pi^B+\ou{A}{B}{C}(\partial_t)\pi^C$ denotes the covariant derivative, being the infinitesimal version of the bulk holonomies $g\in SL(2,\C)$. Putting all the pieces together, the zeroth order cancels and we find that the Lagrangian starts at linear order in epsilon:
\begin{equation}
S_{w_i}^{\mathrm{top}}=\frac{\varepsilon}{2}\Big(\omega_A(t) D_{\partial_t}\pi^A(t)+\pi_A(t) D_{\partial_t}\omega^A(t)\Big)+\mathcal{O}(\varepsilon^2)+\CC\label{wactn3}
\end{equation}
Summing the contributions from all infinitesimal wedges $w_1, w_2,\dots, w_N$ and taking the limit $N\rightarrow\infty$ we are left with a line integral:
\begin{equation}
S_{w}^{\mathrm{top}}=\frac{1}{2}\int_0^1\di t\Big(\omega_A(t) D_{\partial_t}\pi^A(t)+\pi_A(t) D_{\partial_t}\omega^A(t)\Big)+\CC,\label{wactn3}
\end{equation}
This action, being nothing but a covariant symplectic potential, generates trivial equations of motion, which just tell us $\omega$ and $\pi$ are parallel along the edge $e\subset\partial w$:
\begin{equation}
D_{\partial_t}\omega^A=0=D_{\partial_t}\pi^A\label{BFmotion}
\end{equation}
What is more important, concerns the area-matching condition. On each infinitesimal wedge $w_i$ equation \eref{armtch} becomes:
\begin{equation}
C=\epsilon^{AB}\pi_A^{(i+1)}\omega_B^{(i+1)}-\epsilon^{AB}\pi_A^{(i)}\omega_B^{(i)}=E(t+\varepsilon)-E(t)
\end{equation}
Therefore, when taking the continuum limit in the sense of this section, the area-matching condition turns into the conservation law:
\begin{equation}
\dot{E}=\frac{\di}{\di t}(\pi_A\omega^A)=D_{\partial_t}(\pi_A\omega^A)\stackrel{!}{=}0.\label{conslaw}
\end{equation}  
We can now see the area-matching-constraint is satisfied just because the equations of motion \eref{BFmotion} guarantee $\dot{E}=0$ holds for all times.
\subsection{The constrained continuum action on an edge}\noindent
In the following, we will rearrange the sum over wedges to find the contribution to the total action from a single edge $e$. Each edge carries a unit time-like four-vector $n^\alpha$ representing the internal (future pointing) Minkowski normal of the tetrahedron dual to the edge. We have started from a discrete model, where we know this normal only somewhere in the middle of the edge (say at parameter time $t=t_o=\tfrac{1}{2}$). 
Employing local Lorentz invariance we can always put that normal into the canonical gauge, i.e.:
\begin{equation}
n^{A\bar A}(t=t_o=\tfrac{1}{2})=\frac{\I}{\sqrt{2}}\delta^{A\bar A}=n^{A\bar A}_o.\label{tgaugeedge}
\end{equation}
But now that we have a continuous action we need this normal all along the edge. We achieve this by using one of the key assumptions of spinfoam gravity, that states the geometry be locally flat. This implies the normal is covariantly constant along the edge. To be more precise, if we take the edge to be parametrised by our coordinate $t\in[0,1]$, we assume
\begin{equation}
\forall t\in (0,1):D_{\partial_t}n^\alpha=0.\label{nrmevolv}
\end{equation}
Notice the boundary values $0$ and $1$ are excluded here, reflecting the fact that \eref{nrmevolv} holds locally but cannot be achieved all around the spinfoam face.  
In fact a number of edges, say $e_1$, $e_2$, $\dots e_N$, bound a spinfoam face, along each of which we can introduce a continuous time variable $t_1\in(0,1]$, $t_2\in(1,2]$, $\dots t_N\in(N-1,N)$ along the lines of the last section.
Wherever two edges meet, that is on a spinfoam vertex, a discontinuity may arise, measured by the angle $\Xi_i$:
\begin{equation}
\mathrm{ch}(\Xi_i)=-\lim_{\varepsilon\searrow 0}n_\alpha(i-\varepsilon)n^\alpha(i+\varepsilon).\label{dihdrl}
\end{equation}

Let us now come back to the main issue of this section, the action for an edge. At every edge four triangles $\tau_I$ with $I=1,\dots 4$ meet, for each of which we introduce spinors\footnote{In the following we will keep the index $(I)$ only when it is strictly necessary, to further simplify our notation we will also put this index wherever there is \qq{enough} space, e.g. $\pi^A_{(I)}=\epsilon^{AB}\pi^{(I)}_B$.} $(\omega^{(I)},\pi^{(I)})$. The topological action on an edge becomes (after having shifted the boundaries to $0$ and $1$ again):
\begin{equation}
S_{e}^{\mathrm{top}}=\frac{1}{2}\sum_{I=1}^4\int_0^1\di t\Big(\omega_A^{(I)} D_{\partial_t}\pi^A_{(I)}+\pi_A^{(I)} D_{\partial_t}\omega^A_{(I)}\Big)+\CC,\label{eactntop}
\end{equation}  
Before we add the reality conditions \eref{simplspin} to this action, let us first discuss the last constraint missing, i.e. the Gauß law.
We find it from the variation of the edge action \eref{eactntop} with respect to the selfdual connection $\ou{A}{A}{B}$ contracted with the tangent vector $\partial_t=\dot{e}$ of the edge. The resulting $\mathfrak{sl}(2,\C)$-valued Lagrange multiplier
\begin{equation}
\Phi^{AB}(t):=A^{AB}_{e(t)}(\partial_t)\label{lagmult}
\end{equation}
appears linearly in the covariant derivative $D_{\partial_t}\pi^A=\dot{\pi}^a+\ou{\Phi}{A}{B}\pi^B$.
The variation of the action \eref{eactntop} with respect to this Lagrange multiplier leads us to the Gauß constraint
\begin{equation}
G_{AB}=-\frac{1}{2}\sum_{I=1}^4\pi_{(A}^{(I)}\omega_{B)}^{(I)}=\sum_{I=1}^4\Pi_{AB}^{(I)}=\sum_{I=1}^4\Pi^{(I)}_i\uo{\tau}{AB}{i}=0.
\end{equation}
When introducing real and imaginary parts of $2\Pi_i=L_i+\I K_i$ corresponding to boosts and rotations with respect to the standard normal \eref{nrmevolv} the Gauß law turns into:
\begin{equation}
\sum_{I=1}^4 L_i^{(I)}=0=\sum_{I=1}^4 K_i^{(I)}.\label{gsslaw}
\end{equation}
If we now remember the reality conditions imply the combination $K_i+\beta L_i$ vanishes on all triangles, we can see that not all the constraints are independent. If we impose the reality conditions for the three triangles $I=1,2,3$, the Gauß law immediately implies them for the fourth. 

Conversely, we can introduce Lagrange multipliers $z_{(I)}:[0,1]\rightarrow\C$ and $\lambda_{(I)}:[0,1]\rightarrow\R$ that enforce the reality conditions on three triangles only. We thus get for the action on an edge:
\begin{equation}
S_{e}=\frac{1}{2}\sum_{I=1}^4\int_0^1\di t\Big(\omega_A^{(I)} D_{\partial_t}\pi^A_{(I)}+\pi_A^{(I)} D_{\partial_t}\omega^A_{(I)}
-2z_{(I)}F_n[\pi_{(I)},\omega_{(I)}]-\lambda_{(I)}D[\pi_{(I)},\omega_{(I)}]\Big)+\CC,\label{eactn}
\end{equation}
with the gauge-fixing condition
\begin{equation}
z_{(4)}=0=\lambda_{(4)}.\label{zgauge}
\end{equation}
Once we have an action, we should discuss the equations of motion. This involves two steps. To begin with, in section \ref{IIID}, we are going to study the constraints and perform the Dirac algorithm \cite{dirac}. Then we also have to study the evolution equations, and ask for their geometric interpretation, which we are going to do in sections \ref{IIID}, \ref{IIIE} and \ref{IIIF}.
\subsection{Dirac analysis of all constraints}\label{IIID}\noindent
Let us turn to whether the equations of motion preserve the constraints. We do this in the Hamiltonian picture. With the Lagrangian \eref{eactn}, linear in the time derivatives, we can immediately find the Hamiltonian, which is itself constrained to vanish. If we introduce the \emph{primary Hamiltonian}
\begin{equation}
H^\prime[\pi,\omega](t)=z(t)F_{n(t)}[\pi,\omega]+\frac{\lambda(t)}{2}D[\pi,\omega]+\CC,\label{primham}
\end{equation}
we can write the evolution equations in the most covariant way possible:
\begin{equation}
D_{\partial_t}\omega^A=\{H^\prime,\omega^A\},\quad D_{\partial_t}\pi^A=\{H^\prime,\pi^A\}.
\end{equation}
The canonical commutation relations are $\{\pi^A,\omega^B\}=\epsilon^{AB}$, and $D_{\partial_t}$ is again the covariant $\mathfrak{sl}(2,\C)$ derivative $D_{\partial_t}\pi^A=\partial_t{\pi}^A+\ou{\Phi}{A}{B}\pi^B$, with $\ou{\Phi}{A}{B}$ being the selfdual connection contracted with the tangent vector of the edge---just as defined in \eref{lagmult}. 
\newline
\newline
\noindent To prove that the Hamiltonian vector-field preserves the constraints we discuss each of them separately.
\paragraph*{(i) stability of the area-matching constraint $\dot{E}=0$, and of $D=0$.} The area-matching constraint $\dot{E}=0$ guarantees the area of a triangle is the same seen from all tetrahedra it belongs to. The Hamiltonian vector field of $E=\pi_A\omega^A$ acts as follows:
\begin{equation}
\{E,\pi^A\}=-\pi^A,\quad\{E,\omega^A\}=\omega^A,\quad\{E,\bar\pi^{\bar A}\}=0=\{E,\bar\omega^{\bar A}\}.
\end{equation}
We thus easily get
\begin{equation}
\dot{E}=D_{\partial_t}E=\{H^\prime,E\}=z(t)F-\bar{z}(t)\bar{F}\propto 0,\label{Econs}
\end{equation}
where $\propto$ means equality up to constraints. Since $D=\I E/(\beta+\I)+\CC$, hence linear in $E$,
equation \eref{Econs} also implies that the reality condition $D=0$ holds for all times:  
\begin{equation}
\dot{D}=\{H^\prime,D\}\propto 0,
\end{equation}
Therefore, the Hamiltonian time evolution along a spinfoam edge preserves both the area-matching constraint and the Lorentz invariant part $D=0$ of the simplicity constraints $K_i+\beta L_i=0$.
\paragraph*{(ii) stability of $F_n=0$.} Before we explore under which conditions our primary Hamiltonian \eref{primham} is compatible with the constraint $F_n=0$, let us first recall \cite{komplexspinors} all solutions of the reality conditions $F_n=0=D$. They are parametrised by a real number $J\neq 0$, and tell us the momentum $\pi$ is proportional to $\bar\omega$. We find, in fact,
\begin{equation}
\pi_A=-\I\sqrt{2}(\beta+\I)J\frac{n_{A\bar A}\bar{\omega}^{\bar A}}{\|\omega\|^2_n},\label{gensol}
\end{equation}
with the $SU(2)$ norm $\|\omega\|_n^2=-\I\sqrt{2}n_{A\bar A}\omega^A\bar\omega^{\bar A}$. Notice that we can always assume $J>0$. We have mentioned, in the lines shortly above equation \eref{confsym}, that there is a discrete symmetry simultaneously exchanging all $\pi$ and $\omega$ spinors. Since
\begin{equation}
J=\frac{\pi_A\omega^A}{\beta+\I},\label{Jinterpret}
\end{equation}
a transformation exchanging $\pi$ and $\omega$, maps $J$ into $-J$, hence $J>0$ without loss of generality. 

The quantity $J$ parametrising the solutions of the reality conditions also has a clean geometrical interpretation. It measures the area $A[\tau]$ of the triangle $\tau$ under consideration. A short calculation gives the precise relation:
\begin{equation}
A^2[\tau]=\Sigma^i[\tau]\Sigma_i[\tau]=-2\Sigma^{AB}\Sigma_{AB}=\Big(\frac{\beta\ellp^2 J}{\hbar}\Big)^2\label{triar}
\end{equation}

We are now ready to come back to our original problem, and show how the Hamiltonian can preserve the reality conditions.
Since the normal is covariantly constant, we get for the time evolution of $F_n=0$ the equation:
\begin{equation}
\dot{F}_n=\frac{\di}{\di t}\big(n^{A\bar A}\pi_A\bar{\omega}_{\bar A}\big)=D_{\partial_t}\big(n^{A\bar A}\pi_A\bar{\omega}_{\bar A}\big)=\big\{{H}^\prime,F_n\big\}\propto\bar{z}(t)\big\{\bar{F}_n,F_n\big\}
\end{equation}
We calculate the missing Poisson bracket in the gauge where $n^\alpha=n^\alpha_o$ and find:
\begin{align}
\nonumber\big\{\bar{F}_{n_o},F_{n_o}\big\}&=\frac{1}{2}\delta_{A\bar A}\delta_{B\bar B}\big\{\bar{\pi}^{\bar A}\omega^A,\pi^B\bar{\omega}^{\bar B}\big\}
=\frac{1}{2}\delta_{A\bar A}\delta_{B\bar B}\Big[\bar{\epsilon}^{\bar A\bar B}\omega^A\pi^B+\epsilon^{AB}\bar{\pi}^{\bar A}\bar{\omega}^{\bar B}\Big]=\\
&=-\frac{1}{2}\big(\pi_A\omega^A-\CC\big)=-\I J
\end{align}
The result being manifestly $SL(2,\C)$ invariant we can conclude that
\begin{equation}
\dot{F}_n\propto-\I\bar{z}(t) J.
\end{equation}
We have assumed the area of the triangle does not vanish, hence $J\neq 0$. This implies the Hamiltonian flow preserves the constraint $F_n=0$ only if we put the Lagrange multiplier $z(t)$ to zero. Reinserting this restriction on the Lagrange multiplier into the primary Hamiltonian \eref{primham} we get the secondary Hamiltonian
\begin{equation}
H^{\prime\prime}=\lambda(t)D[\pi,\omega]\label{secham}
\end{equation} 
\paragraph*{(iii) stability of the Gauß law.} 
The secondary Hamiltonian \eref{secham} generates the edge-evolution compatible with the simplicity constraints for one pair of spinors. There are, however, four of these pairs per edge---one twistor $Z=(\bar\pi_{\bar A},\omega^A)$ for each adjacent triangle. The Gauß law is an example of an observable depending on all of them. Its time evolution is governed by the physical Hamiltonian, which is the sum over the secondary Hamiltonians \eref{secham} of all four triangles:
\begin{equation}
H^{\mathrm{phys}}=\sum_{I=1}^4\lambda_{(I)}(t)D[\pi_{(I)},\omega_{(I)}].
\end{equation}
The Hamiltonian has this simple form, just because the action for an edge \eref{eactn} splits into a sum over adjacent triangles, without any \qq{interaction-terms} appearing.
Since the Hamiltonian vector field of the constraint $D=0$ acts as
\begin{equation}
\mathfrak{X}_D[\omega^A]=\{D,\omega^A\}=\frac{\I}{\beta+\I}\omega^A,\quad\mathfrak{X}_D[\pi^A]=\{D,\pi^A\}=-\frac{\I}{\beta+\I}\pi^A\label{dtrafos}
\end{equation}
we immediately get for any choice of $\lambda$, that the Gauß constraint is covariantly preserved:
\begin{equation}
D_{\partial_t}G_{AB}=\big\{H^{\mathrm{phys}},G_{AB}\big\}=0.
\end{equation}
The partial derivative, on the other hand, vanishes weakly:
\begin{equation}
\frac{\di}{\di t}G_{AB}\propto 0.
\end{equation}
which follows from the commutation relations of the Lorentz algebra:
\begin{equation}
\{L_i,L_j\}=-\uo{\epsilon}{ij}{l}L_l,\quad
\{L_i,K_j\}=-\uo{\epsilon}{ij}{l}K_l,\quad
\{K_i,K_j\}=+\uo{\epsilon}{ij}{l}L_l.
\end{equation}
\paragraph*{First- and second-class constraints.}  We got the constraint equations on an edge by varying the Lagrange multipliers $\Phi\in\sl(2,\C)$, $z_{(I)}\in\C$ and $\lambda_{(I)}\in\R$ in the action \eref{eactn}. If we want to quantise the theory we have to compute the constraint algebra and identify first-class and second-class constraints therein. The set of constraints consists of both the rotational and boost part of the Gauß law, together with simplicity constraints on the triangles $I=1,2,3$. The simplicity constraints $K_i+\beta L_i=0$ are imposed on three triangles only, but the Lorentzian Gauß law \eref{gsslaw} implies them on the fourth. These constraints can be rearranged to treat all triangles equally. We can then impose just the rotational part of the Gauß law, and require the simplicity constraints \eref{simplspin} on all four triangles. This rearrangement leads us to the following system of constraints:
\begin{equation}
G_i^{\mathrm{rot}}=\sum_{I=1}^4L_i[\pi_{(I)},\omega_{(I)}]\stackrel{!}{=}0,\quad\forall I=1,\dots,4: D_{(I)}\equiv D[\pi_{(I)},\omega_{(I)}]\stackrel{!}{=}0\stackrel{!}{=}F_{n_o}[\pi_{(I)},\omega_{(I)}]\equiv F_{(I)},\label{setcons}
\end{equation}
Notice, that in a general gauge, where the time normal does not assume the canonical form $n^\alpha=n^\alpha_o$, we must boost the constraints into the direction of $n^\alpha$. We would then work with the constraints
\begin{equation}
G_{i(n)}^{\mathrm{rot}}:=\mathrm{exp}\big(\mathfrak{X}_{G_i^{\mathrm{boost}}}\eta^i\big)G_i^{\mathrm{rot}},\quad\text{and}\quad
F_n=\mathrm{exp}\big(\mathfrak{X}_{G_i^{\mathrm{boost}}}\eta^i\big)F_{n_o}\label{boostng}
\end{equation}
instead. Here, $\mathfrak{X}_{G_i^{\mathrm{boost}}}=\{G_i^{\mathrm{boost}},\cdot\}$ denotes the Hamiltonian vector-field of the boost part of the Gauß law and the generic normal $n$ has been parametrised as:
\begin{equation}
(n^0,n^i)=\Big(\mathrm{ch}(|\eta|),\;\mathrm{sh}(|\eta|)\frac{\eta^i}{|\eta|}\Big),\quad \text{where:}\; |\eta|=\sqrt{\delta_{ij}\eta^i\eta^j}.
\end{equation}
All of the constraints \eref{setcons} are preserved by the physical Hamiltonian generating the time evolution along an edge, e.g. ${\dot{G}}{}_{i(n)}^{\mathrm{rot}}=\{{H}^{\mathrm{phys}},{G}_{i(n)}^{\mathrm{rot}}\}\propto 0.$
To identify first- and second-class constraints within this set, we have to study their mutual Poisson brackets. We find: 
\begin{subequations}
\begin{align}
&\{G_{i}^{\mathrm{rot}},G_{j}^{\mathrm{rot}}\}=-\uo{\epsilon}{ij}{l}G_{l(n)}^{\mathrm{rot}},\quad
\{G_{i}^{\mathrm{rot}},D_{(I)}\}=0,\quad
\{G_{i}^{\mathrm{rot}},F_{(I)}\}=0=\{G_i^{\mathrm{rot}},\bar{F}_{(I)}\}=\\
&\{D_{(I)},F_{(J)}\}=-\frac{2\I\beta}{\beta^2+1}\delta_{IJ}F_{(I)},\quad\{D_{(I)},\bar{F}_{(J)}\}=
\frac{2\I\beta}{\beta^2+1}\delta_{IJ}\bar{F}_{(I)}\\
&\{F_{(I)},\bar{F}_{(J)}\}=\I\delta_{IJ}\Im(\pi_A^{(I)}\omega^A_{(I)})=\I\delta_{IJ}\Im(E_{(I)}).
\end{align}\label{algbr}
\end{subequations}
The set of first class constraints consists of the Gauß law, attached to each edge, together with the Lorentz invariant simplicity constraint $D=0$, attached to each triangle. The constraint $F_n=0$ is second class and generates, an additional $\su(2)$ algebra. This becomes more explicit once we define the ladder operators $J_\pm$ together with the generator $J_3$:
\begin{equation}
J_-=J_1-\I J_2=:-\sqrt{2}\bar{F}_{n_o},\quad J_+=J_1+\I J_2:=-\sqrt{2}{F}_{n_o},\quad J_3:=\Im(E),\label{hiddensu2}
\end{equation}
with the Poisson bracket of the rotation group:
\begin{equation}
\{J_i,J_k\}=-\uo{\epsilon}{ij}{k}J_k.
\end{equation}
In our case $J_-$ and $J_+$ are constrained to vanish, while $J_3\neq 0$, reflects the fact that the constraints form a second-class system. A last comment on the time-gauge \eref{tgaugeedge}: if we want to relax this condition, little will happen, the constraints get boosted \eref{boostng} but the structure constants appearing in the constraint algebra \eref{algbr} remain the same. 
\subsection{Solving the equations of motion for the spinors}\label{IIIE}\noindent
In this section we will solve the equations of motion for the spinors. For any triangle adjacent to the edge it is the physical Hamiltonian \eref{secham} that generates the time evolution of the corresponding spinors:
\begin{equation}
D_{\partial_t}{\omega}^A=\{H^{\prime\prime},\omega^A\},\quad D_{\partial_t}{\pi}^A=\{H^{\prime\prime},\pi^A\}.
\end{equation}
We thus get the following equations of motion:
\begin{subequations}
\begin{align}
\big(D_{\partial_t}{\omega}^A\big)(t)&=\dot{\omega}^A(t)+\ou{\Phi}{A}{B}(t)\omega^{B}(t)=\lambda(t)\big\{D,\omega^A\big\}_t=
+\frac{\I}{\beta+\I}\lambda(t)\omega^A(t),\\
\big(D_{\partial_t}{\pi}^A\big)(t)&=\dot{\pi}^A(t)+\ou{\Phi}{A}{B}(t)\pi^{B}(t)=\lambda(t)\big\{D,\pi^A\big\}_t
=-\frac{\I}{\beta+\I}\lambda(t)\pi^A(t),
\end{align}\label{spinevlv}
\end{subequations}
with $\ou{\Phi}{A}{B}$ defined in \eref{lagmult}. We introduce the parallel transport between time $t$ and $t^\prime$ along the edge
\begin{equation}
U(t,t^\prime)=\mathrm{Pexp}\big(-\int_t^{t^\prime}\di s\Phi(s)\big)\in SL(2,\C),\label{edgehol}
\end{equation}
and use it to write down the general solution of the equations of motion. We get
\begin{subequations}
\begin{align}
\omega^A(t)&=\exp\Big[{+\frac{\I}{\beta+\I}\int_0^t\di s\lambda (s)}\Big]\ou{U}{A}{B}(0,t)\omega^A(0),\\
\pi^A(t)&=\exp\Big[{-\frac{\I}{\beta+\I}\int_0^t\di s\lambda (s)}\Big]\ou{U}{A}{B}(0,t)\pi^A(0).
\end{align}\label{spinsol}
\end{subequations}
\paragraph*{Closing the loop.} In \eref{spinsol} we found the solution of the equations of motion for the spinors on an edge. The spinors represent the flux \eref{gravfluxdef} through a triangle seen from the frame of the tetrahedron dual to the edge. But the triangle belongs to many tetrahedra, hence many edges.   
These edges, $e_1$, $e_2$, $\dots e_N$, bound the spinfoam face dual to the triangle. Just as we have done above, we can introduce a continuum time variable $t_1\in(0,1]$, $t_2\in(1,2]$, $\dots t_N\in (N-1,N)$ for each of these edges, and study the field of spinors $Z=(\bar{\pi}_{\bar A},\omega^A):[0,N)\ni t\mapsto \C^2{}^\ast\otimes\C^2$ along the edge. This field describes the triangle dual to the spinfoam face in the frame of the various edges. To guarantee $Z(t)$ describes, for all $t\in[0,N)$, the same triangle we need boundary conditions:
\begin{subequations}
\begin{equation}
\forall i\in\{1,\dots, N-1\}:\lim_{\varepsilon\searrow 0}\omega^A(i+\varepsilon)=\lim_{\varepsilon\searrow 0}\omega^A(i-\varepsilon), \quad \lim_{\varepsilon\searrow 0}\pi^A(i+\varepsilon)=\lim_{\varepsilon\searrow 0}\pi^A(i-\varepsilon),\label{bound1}
\end{equation} 
and also
\begin{equation}
\lim_{\varepsilon\searrow 0}\omega^A(\varepsilon)=\lim_{\varepsilon\searrow 0}\omega^A(N-\varepsilon), \quad \lim_{\varepsilon\searrow 0}\pi^A(\varepsilon)=\lim_{\varepsilon\searrow 0}\pi^A(N-\varepsilon).
\end{equation}\label{bound2}
\end{subequations}\label{bnd}
Using the evolution equations \eref{spinsol} we find that the boundary conditions (\ref{bound2}) turn into a constraint on the holonomy once we close the loop around the spinfoam face. This happens as follows: We invert \eref{spinsol} and solve it for the holonomy $U(0,N)$ in terms of the quadruple of spinors $(Z(0),Z(N))$. Inserting the boundary conditions we get:
\begin{equation}
\ou{U}{A}{B}(0,N)=\mathrm{Pexp}\ou{\big(-\int_0^N\di s\Phi(s)\big)}{A}{B}=(\pi\omega)^{-1}\Big(\E^{-\frac{\I}{\beta+\I}\Lambda}\omega^A(0)\pi_B(0)-\E^{+\frac{\I}{\beta+\I}\Lambda}\omega^A(0)\pi_B(0)\Big)
\label{reggehol}
\end{equation}
where $\pi\omega=\pi_A\omega^A$, and we have introduced the quantity
\begin{equation}
\Lambda=\int_0^N\di t\lambda(t),\label{deffangle}
\end{equation}
which we can write equally well as
\begin{equation}
\exp\Big(-\frac{2\Lambda}{\beta^2+1}\Big)=\frac{\|U(0,N)\omega(0)\|^2_n}{\|\omega(0)\|^2_n}.
\end{equation}
Equation \eref{reggehol} is an interesting result. First of all it tells us that the holonomy around a spinfoam face cannot be a generic $SL(2,\C)$ element but preserves the flux through the triangle dual to the spinfoam face, i.e.:
\begin{equation}
\ou{U}{A}{C}(0,N)\ou{U}{B}{D}(0,N)\omega^{(C}(0)\pi^{D)}(0)=\omega^{(A}(0)\pi^{B)}(0)
\end{equation}
The same constraint also appears in Regge calculus \cite{Reggecalc, Reggebonzom, Reggebarrett, ReggeDittrichSimone, Reggebrewin}, but there is a major difference. In Regge calculus the holonomy \eref{reggehol} is further constrained to be a pure boost, here it is neither a boost, nor a rotation, but a four-screw, i.e. a combination of a rotation and a boost in the direction of the rotation axis.

In the next two sections we will further delve in the geometry of the spinfoam face and prove that $\lambda$ is a measure of both extrinsic and intrinsic curvature.
\subsection{Extrinsic curvature}\label{IIIF}\noindent
We are now going to calculate the extrinsic curvature smeared along a link connecting two adjacent tetrahedra. This will give us a better understanding of the Lagrange multiplier $\lambda$ appearing in the action \eref{eactn}. We will indeed prove, that it is a measure of the extrinsic curvature smeared along a link.  

Let us consider first two points labelled by coordinates $t$ and $t^\prime$ on the boundary of the spinfoam face. Take the holonomy $h(t,t^\prime)$ along the link connecting the two respective tetrahedra sitting at $t$ and $t^\prime$, we use \eref{holdef2} and thus have:
\begin{equation}
\ou{h}{A}{B}(t,t^\prime)=\frac{\utilde{\omega}^A\pi_B-\utilde{\pi}^A\omega_B}{\sqrt{\pi\omega}\sqrt{\utilde{\pi}\utilde{\omega}}},
\end{equation}
where we have introduced the abbreviation
\begin{equation}
(\utilde{\omega},\utilde{\pi},\omega,\pi)=(\omega(t^\prime),\pi(t^\prime),\omega(t),\pi(t)).
\end{equation}
At this point let us stress again that links and edges have to be carefully distinguished. Edges enter the bulk of 4-simplices, whereas links belong to the three-dimensional boundary of the 4-simplex, see figure \ref{simplx}.

The two tetrahedra are embedded into the four dimensional manifold with normals $n=n(t)$ and $\utilde{n}=n(t^\prime)$. The extrinsic curvature smeared over the link between the two is measured by the angle \cite{twistintegrals}:
\begin{equation}
\begin{split}
\mathrm{ch}(\Xi(t,t^\prime))&=-{\utilde{n}}_{A\bar{A}}\ou{h}{A}{B}(t,t^\prime)\ou{\bar{h}}{\bar{A}}{\bar{B}}(t,t^\prime)n^{B\bar B}=-\frac{1}{|\pi\omega|^2}
\utilde{n}_{A\bar A}
\big(\utilde{\omega}^A\pi_B-\utilde{\pi}^A\omega_B\big)
\big(\utilde{\bar\omega}^{\bar A}\bar\pi_{\bar B}-\utilde{\bar\pi}^{\bar A}\bar\omega_{\bar B}\big)
n^{B\bar B}=\\
&=\frac{1}{2}\Big(\frac{\|\omega\|^2_n}{\|\utilde{\omega}\|^2_{\utilde{n}}}+\frac{\|\utilde{\omega}\|^2_n}{\|\omega\|^2_{\utilde{n}}}\Big).
\end{split}
\end{equation}
This equation gives the angle up to a sign, we remove the remaining ambiguity, just as in reference \cite{asymptcs}, by defining
\begin{equation}
\E^{\Xi(t,t^\prime)}:=\frac{\|\omega(t)\|^2_{n(t)}}{\|\omega(t^\prime)\|^2_{n(t^\prime)}}.\label{Xidef}
\end{equation}
There are now two important cases to distinguish. In the first one, $t$ and $t^\prime$ lie on the same edge. The normal is parallel along the edge, hence transported by the holonomy according to
\begin{equation}
n^{A\bar A}(t^\prime)=\ou{U}{A}{B}(t,t^\prime)\ou{\bar U}{\bar A}{\bar B}(t,t^\prime)n^{B\bar B}(t).\label{normtrans}
\end{equation}
With this equation the normals cancel from the definition of the angle \eref{Xidef}, and we find the following.
\begin{equation}
\text{If $t,t^\prime$ belong to the same edge:}\quad\E^{\Xi(t,t^\prime)}=\frac{\|\omega(t)\|^2_{n(t)}}{\|\omega(t^\prime)\|^2_{n(t^\prime)}}=\E^{-\frac{2}{\beta^2+1}\int_t^{t^\prime}\di s\lambda (s)}.
\end{equation}
In the second case $t$ and $t^\prime$ belong to neighbouring tetrahedra, with normals $n(t)$ and $n(t^\prime)$ to be distinguished. Assume the two tetrahedra meet at the $i$-th vertex, that is at coordinate value $t=i$. We thus get
\begin{equation}
\E^{\Xi(t,t^\prime)}=\E^{-\frac{2}{\beta^2+1}\int_t^{t^\prime}\di s\lambda(s)}\frac{\|U(t,i)^{-1}\omega(i)\|^2_{n(t)}}{\|U(i,t^\prime)\omega(i)\|^2_{n(t^\prime)}}.
\end{equation}
In the middle of each edge we have chosen time-gauge \eref{tgaugeedge}, hence:
\begin{equation}
\forall i:n^{A\bar A}(\tfrac{2i+1}{2})=\frac{\I}{\sqrt{2}}\delta^{A\bar A}.\label{tgaugemany}
\end{equation}
We compute the angle between adjacent tetrahedra, as introduced in \eref{dihdrl}, and get: 
\begin{equation}
\E^{\Xi_{i}}:=\lim_{\varepsilon\searrow 0}\E^{\Xi(i-\varepsilon,i+\varepsilon)}=
\frac{\|U(i-\varepsilon,i)^{-1}\omega(i)\|^2_{n(i-\varepsilon)}}{\|U(i,i+\varepsilon)\omega(i)\|^2_{n(i+\varepsilon)}}
=\frac{\|g_{e_i^{\mathrm{target}}}\omega(i)\|^2}{\|g_{e_{i+1}}^{\mathrm{source}}\omega(i)\|^2},
\label{dihedrlangl}
\end{equation}
with the norm $\|\omega\|^2=\delta_{A\bar A}\omega^A\bar\omega^{\bar A}$, and the abbreviations:
\begin{equation}
g_{e_i}^{\mathrm{target}}=U(i,\tfrac{2i-1}{2}),\quad g_{e_i}^{\mathrm{source}}=U(i-1,\tfrac{2i-1}{2}).\label{sourcetarget}
\end{equation} 
These goup elements belong to the final and initial point of the edges; $g_{e_i}^{\mathrm{target}}$, for example, is the $SL(2,\C)$ holonomy along the $i$-th half-edge going from the vertex $v_i$ towards the center of the edge at parameter time $t=\tfrac{2i-1}{2}$. These bulk holonomies play an important role in the asymptotic analysis of the spinfoam amplitude \cite{asymptcs, Hanasymptotics}, which is why we have introduced them here explicitly.

Do the $\Xi$-angles just considered define proper observables? There are two gauge symmetries to take care off, the $SU(2)$ transformations generated by $G_i^{\mathrm{rot}}$ \eref{setcons} (or rather $G_{i(n)}^{\mathrm{rot}}$ for the more general case), and the scaling transformations generated by the Hamiltonian vector-field \eref{dtrafos} of $D$. Since the $SU(2)$ norm is, by definition, rotational invariant, and the angles are a function of those, they are certainly $SU(2)$ invariant too. But $\Xi(t,t^\prime)$ transforms non-trivially under $D$. We have in fact:
\begin{equation}
\|\omega\|^2_n\mapsto\exp(\varepsilon\mathfrak{X}_D)\big[\|\omega\|^2_n\big]=\E^{\frac{2}{\beta^2+1}\varepsilon}\|\omega\|^2_n.
\end{equation}
Since $\varepsilon$ may locally be an arbitrary continuous function of $t$, the gauge transformation shifts the integral over the Lagrange multiplier to a new value:
\begin{equation}
\int_t^{t^\prime}\di s\lambda(s)\mapsto\varepsilon(t^\prime)-\varepsilon(t)+\int_t^{t^\prime}\di s\lambda(s),\quad\text{thus:}\;\lambda\mapsto\dot{\varepsilon}+\lambda.
\end{equation}
We see $\Xi(t,t^\prime)$ is generally not $D$-invariant and does not define a proper observable. Nevertheless there is a gauge invariant quantity, that we can build out of $\lambda$. The overall angle, as defined in \eref{deffangle} is an observable. This is true, simply because we are working with periodic boundary conditions \eref{bound2} that require periodicity $\varepsilon(0)=\varepsilon(N)$ of the gauge parameter.

We can make the gauge invariance of $\Lambda$ even more obvious. Notice first that any transformation generated by $D$ cannot change the angles (\ref{dihdrl}, \ref{dihedrlangl}) between adjacent tetrahedra. Consider next the boundary conditions (\ref{bound2}). They imply all angles $\Xi(\tfrac{2i-1}{2},\tfrac{2i+1}{2})$ sum up to zero when going around a spinfoam face. This means
\begin{equation}
1=\E^{\sum_{i=1}^N\Xi_i}\E^{-\frac{2}{\beta^2+1}\int_0^N\di t\lambda(t)},\quad\text{thus}\quad\sum_{i=1}^N\Xi_i= \frac{2}{\beta^2+1}\Lambda.\label{lambdainterpret}
\end{equation}
The last identity gives $\Lambda$ in terms of the angles $\Xi_i$ between the normals of adjacent tetrahedra. In the next section we prove this quantity is proportional to the curvature tensor smeared over the spinfoam face, revealing a close analogy with Regge calculus. This proportionality will be exact and not an approximation.
\subsection{Intrinsic curvature}\label{IIIG}\noindent
The previous sections revealed a Hamiltonian generating the time evolution along a spinfoam edge. We have seen this Hamiltonian preserves the constraint equations---the Gauß law together with the simplicity constraints---once the Lagrange multiplier in front of the second-class constraint $F_n=0$ vanishes. Both Gauß's law and the simplicity constraints have a well explored physical interpretation, they guarantee all triangles represent spatial planes in internal Minkowski space that close to form a tetrahedron \cite{polyhdr, bianchisommer}.  
Knowing the geometric interpretation of the constraints, what do the evolution equations tell us? Do they also have a clean physical interpretation? In this section we will explore this questions, and show that the equations of motion for the spinors probe the curvature smeared along a wedge. For this we need some preparations. Let us first study how the holonomy changes under variations of the path.  
 
Be $\gamma_\varepsilon:s\in[0,1]\mapsto\gamma_\varepsilon(s)\in\mathcal{S}$ a $\varepsilon$-parameter family of paths, piecewise differentiable in both $\varepsilon$ and $s$. We can now take two derivatives obtaining the tangent vector $\gamma^\prime_\varepsilon(s)=\frac{\di}{\di s}\gamma_\varepsilon(s)\in T_{\gamma_\varepsilon(s)}\mathcal{S}$ and the variation $\delta{\gamma}_\varepsilon(s)=\frac{\di}{\di \varepsilon}\gamma_\varepsilon(s)\in T_{\gamma_\varepsilon(s)}\mathcal{S}$. For $\varepsilon=0$ we write, e.g. $\delta\gamma(s):=\delta\gamma_{\varepsilon=0}(s)$. From the defining differential equation of the holonomy, i.e.
\begin{equation}
\frac{\di}{\di s}h_{\gamma_{\varepsilon}(s)}=-A_{\gamma_{\varepsilon}(s)}(\gamma^\prime_\varepsilon)h_{\gamma_{\varepsilon}(s)},\label{defeq}
\end{equation}  
we can get the variation of the parallel transport at $\varepsilon=0$. We just need to differentiate equation \eref{defeq} with respect to $\varepsilon$, multiply everything by $h_{\gamma_{\varepsilon}(s)}^{-1}$ and integrate the resulting quantity against $\int_0^1\di s$. Performing a partial integration we then get the variation of the holonomy
\begin{equation}
\frac{\di}{\di\varepsilon}\Big|_{\varepsilon=0}h_{\gamma_\varepsilon(1)}=-A_{\gamma(1)}(\delta\gamma)h_{\gamma(1)}+h_{\gamma(1)} A_{\gamma(0)}(\delta\gamma)+
\int_0^1\di sh_{\gamma(1)} h_{\gamma(s)}^{-1}F_{\gamma(s)}(\gamma^\prime,\delta\gamma)h_{\gamma(t)}.\label{varhol}
\end{equation}
\begin{figure}
     \centering
     \includegraphics[width= 0.30\textwidth]{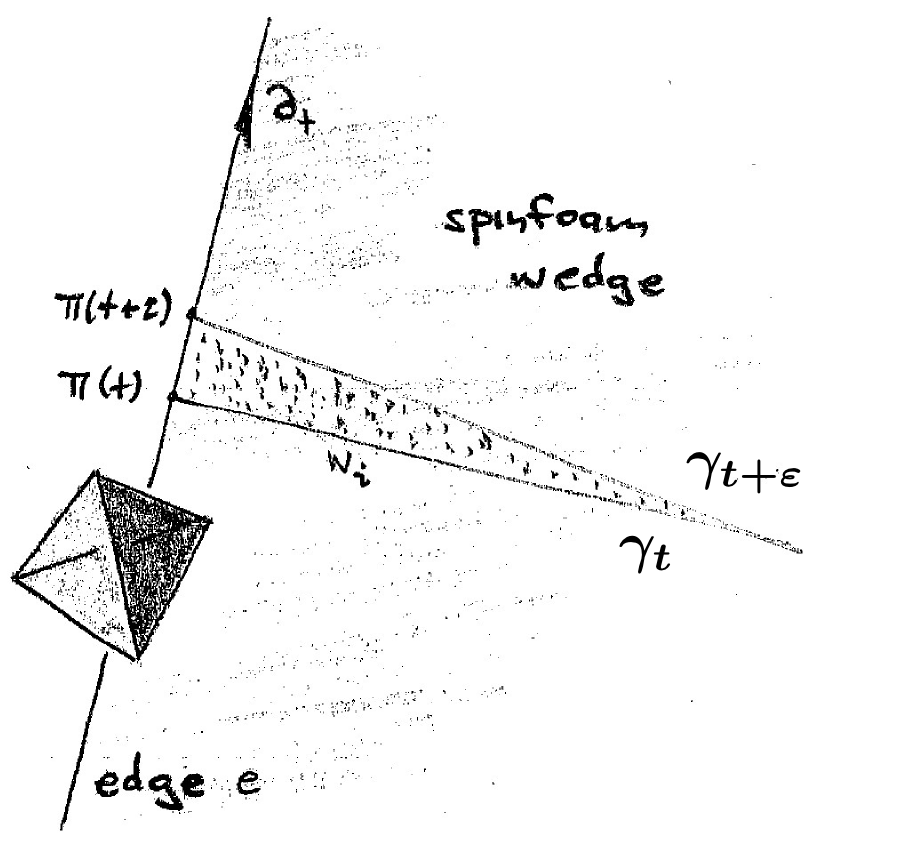}
     \caption{Going from $t$ to $t+\varepsilon$ we can probe an infinitesimal wedge, the boundary of which has two parts. The first part belongs to the edge and has a tangent vector $\partial_t$. The second part (the triangular line in the picture) is a link inside the wedge, itself splitting into two parts. The upper part in the picture we call $\gamma_{t+\varepsilon}$, while the lower part is $\gamma_{t}$, putting them together determines $\pi(t+\varepsilon)$: The spinor $\pi(t+\varepsilon)$ is the parallel transport of $\pi(t)$ along the connected link $\gamma_{t+\varepsilon}^{-1}\circ\gamma_t$.}
     \label{inft}
\end{figure}
Let us now see how the equations of motion for the spinors define such a variation. In our original continuum limit, discussed in section \ref{sectionIIB}, the quadruple $(\pi(t+\varepsilon), \omega(t+\varepsilon), \pi(t), \omega(t))$ probe the $SL(2,\C)$ holonomy-flux variables on an infinitesimal wedge $w_i$. The spinors parametrise the holonomy along the link $\gamma_{t+\varepsilon}^{-1}\circ\gamma_{t}$ connecting the tetrahedra at $t$ and $t+\varepsilon$ by:
\begin{equation}
\pi^A(t+\varepsilon)=\ou{h_{\gamma_{t+\varepsilon}(1)}^{-1}h_{\gamma_{t}(1)}}{A}{B}\pi^{B},\quad\omega^A(t+\varepsilon)=\ou{h_{\gamma_{t+\varepsilon}(1)}^{-1}h_{\gamma_{t}(1)}}{A}{B}\omega^{B}.\label{infwedge}
\end{equation} 
This is just equation \eref{holdef} written in terms of the continuous variables on an edge.
The underlying path $\gamma_{t+\varepsilon}^{-1}\circ\gamma_{t}$ is defined as follows: Three lines bound the infinitesimal wedge, the first goes along the edge, from $t$ towards $t+\varepsilon$. The second part is $\gamma_{t}$ entering the spinfoam face starting at time $t$. The last part $\gamma_{t+\varepsilon}^{-1}$ closes the loop; it goes from inside the spinfoam face towards the edge at $t+\varepsilon$. Figure \ref{inft} gives an illustration of the geometry. Using equation \eref{varhol} we can now take the covariant derivative of \eref{infwedge} to find:
\begin{equation}
D_{\partial_t}\pi^A(t)=\ou{F}{A}{B}(t)\pi^B(t),\quad D_{\partial_t}\omega^A(t)=\ou{F}{A}{B}(t)\omega^B(t),\label{curvevlv}
\end{equation}
where we have introduced the curvature smeared over an infinitesimal wedge. More explicitly
\begin{equation}
\ou{F}{A}{B}(t)=\int_0^1\di s\,\Big[h_{\gamma_t(s)}^{-1}F_{\gamma_t(s)}\Big(\frac{\di}{\di s}\gamma_t(s), \frac{\di}{\di t}\gamma_t(s)\Big)h_{\gamma_t(s)}\Big]^A_{\phantom{A}B}\in\sl(2,\C).
\end{equation}
Notice that equation \eref{curvevlv} has the structure of a deviation equation, with the deviation vector replaced by a spinor. If we now compare \eref{curvevlv} with our equations of motion as they appear in \eref{spinevlv} we can read off the smeared curvature $F(t)$. What we find is that:
\begin{equation}
F^{AB}=\frac{\I}{\beta+\I}\frac{2\lambda}{\pi\omega}\omega^{(A}\pi^{B)}=\frac{2\hbar}{\beta\ellp^2J}\frac{\lambda}{\beta+\I}\Sigma^{AB}.
\label{Fdef}
\end{equation}
We can now go even further and smear the curvature tensor all along the spinfoam face $f$. The curvature having two free indices, this has to be done in a certain frame. We take additional holonomies along the family of paths $\{\gamma_t\}_{t\in (0,N)}$, that map the spinors attached to the boundary of the spinfoam face towards its center---this is the point where all the wedges come together. Indices referring to the frame at the center of the spinfoam face we denote by $A_o, B_o,\dots$. In this frame, the only $t$-dependence of the integrand is in $\lambda(t)$, and we can immediately perform the $t$-integration to arrive at:
\begin{equation}
\int_f F^{A_oB_o}=\frac{1}{2}\int_0^N\di t F^{A_oB_o}(t)=\frac{\Lambda}{\beta+\I}\frac{\Sigma^{A_oB_o}[\tau]}{A[\tau]}.\label{curvtens}
\end{equation}
The factor of one half appears since every infinitesimal wedge has a triangular shape, $A[\tau]$ is the area \eref{triar} of the triangle, while $\Lambda$ denotes the integral  of $\lambda$ along the boundary of the spinfoam face (see \eref{deffangle}).
With \eref{lambdainterpret} we can see this integral is nothing but the sum of the angles between adjacent tetrahedra at all the vertices the triangle belongs to. We could thus say the curvature smeared over a spinfoam face is proportional to the \qq{deficit angle} $\sum_{i=1}^N\Xi_i$ collected when going around a spinfoam face. Although this sounds very much like Regge calculus, there are two subtle differences appearing. First, and most importantly, the curvature smeared over the spinfoam face does not represent a pure boost as in Regge calculus, but instead a four-screw, which is a combination of a rotation and a boost into the direction that the rotation goes around. The relative strength between these two components is measured by the Barbero--Immirzi parameter, which may be an important observation when we ask for the classical role of that parameter.
The second difference is more technical. In Regge calculus, curvature is distributional, and concentrated on the triangles of the simplicial decomposition. Here it is not, but continuously spread over all wedges.  

\subsection{Where could the torsionless condition hide?}\label{IIIH}
\noindent During the last sections we developed a Hamiltonian formalism of the spinfoam dynamics along an edge. The constraint equations must hold for all times, which leads to restrictions on the Lagrange multipliers infront of the second-class constraints. In fact, the multiplier $z$ imposing $F_n=0$ just vanishes. However no secondary constraints appear. This should come as a surprise to us. In the continuum, time evolution preserves the reality conditions only if additional secondary constraints hold true. Together with the  restrictions on the Lagrange multipliers, and the evolution equations for the triad, they force the Lorentz connection to be torsionless, i.e. $T^\alpha=D\wedge e^\alpha=0$.

Let us now ask where the torsionless condition can show up in a discrete theory of gravity. Torsion is a 2-form, which suggests to smear it over the \qq{natural} two-dimensional structures appearing. These are the triangles $\tau$, each of which is bounded by three lines froming the \qq{bones} of the spatial triangulation. With the covariant version of Stoke's theorem the integral over the triangle turns into a sum over the bones $b\in\partial\tau$ bounding the surface:
\begin{equation}
T^\alpha:=D\wedge e^\alpha=0\Rightarrow\sum_{b\in\partial\tau}e^\alpha[b]=0.\label{trsnless}
\end{equation}
Here $e^\alpha[b]$ denotes the tetrad smeared over a bone parallel transported into the frame at the center of the triangle $\tau$. Despite its simple looking from, this equation becomes rather awkward when entering loop gravity. Our elementary building blocks are area-angle variables---fluxes $\Sigma[\tau]$ smeared over triangles $\tau$. We do not have the length-angle variables of the tetrad formalism at our disposal, and we are thus unable to probe equation \eref{trsnless} directly. The tetrads are in fact complicated functions that require invertability of the fluxes---a highly nontrivial condition in a discrete theory of gravity. 

But assuming \eref{trsnless} holds true, we can deduce equations more suitable for area-angle variables. Consider first the covariant exterior derivative of the Plebanski 2-form $\Sigma_{\alpha\beta}=e_\alpha\wedge e_\beta$. This is a 3-form constrained to vanish due to \eref{trsnless}.
We can integrate this 3-form over any tetrahedron, and obtain---again using the non-Abelian Stoke's theorem---Gauß's law:
\begin{equation}
D\wedge (e^\alpha\wedge e^\beta)=D\wedge\Sigma^{\alpha\beta}=0\Rightarrow\sum_{\tau\in\partial\mathcal{T}}\Sigma^{\alpha\beta}[\tau]=0.\label{gausstors}
\end{equation}
For any tetrahedron, the sum of the fluxes through the bounding triangles must vanish (with the fluxes parallel transported into the center of the tetrahedron). This is just the Gauß constraint, that we have already found in \eref{gsslaw}, and therefore parts of the torsionless condition are already satisfied. We can play this trick one more time, arriving at another torsional constraint.

The vanishing of torsion implies the exterior covariant derivative of the volume 3-form $e_\mu\wedge e_\nu\wedge e_\rho$ must vanish. This defines a 4-form, the integral of which must vanish for any four-dimensional region. We take a 4-simplex surrounding a vertex $v$; it is bounded by tetrahedra $\mathcal{T}\in\partial v$ equipped with normals $n^\alpha[\mathcal{T}]$ in the frame of the center of the 4-simplex. Assume all normals are future oriented, and let $\varepsilon[\mathcal{T}]\in\{-1,1\}$ be the sign needed for the vector $\varepsilon[\mathcal{T}]n^\alpha[\mathcal{T}]$ to point outwards the 4-simplex $v$. In the discrete theory, the integral of the covariant exterior derivative of the volume 3-form turns into a sum over the tetrahedra bounding the integration domain: 
\begin{equation}
-\frac{1}{3!}D\wedge (\epsilon_{\alpha\mu\nu\rho}e^\mu\wedge e^\nu\wedge e^\rho)=D\wedge(n_\alpha d^3\mathrm{vol}_n)=0\Rightarrow\sum_{\mathcal{T}\in\partial v}\varepsilon[\mathcal{T}]n_\alpha[\mathcal{T}]\,{}^3\mathrm{vol}[\mathcal{T}]=0.\label{torscons}
\end{equation}
Here ${}^3\mathrm{vol}[\mathcal{T}]$ denotes the volume of the tetrahedron $\mathcal{T}$, a quantity that we can write fully in terms of fluxes:
\begin{equation}
{}^3\mathrm{vol}[\mathcal{T}]=\frac{\sqrt{2}}{3}\sqrt{\big|\epsilon_{ijk}\Sigma^i[\tau_1]\Sigma^j[\tau_2]\Sigma^k[\tau_3]\big|}.\label{3vol}
\end{equation}
Therefore, the torsional constraint \eref{torscons} fits well into the area-angle calculus of spinfoam gravity. Reference \cite{polyhdr} has already discussed this constraint, what is new here, is the torsional interpretation we gave for it.

Let us also stress, that the additional constraint \eref{torscons} may actually impose another condition often ignored in loop gravity. This is the so-called volume constraint \cite{flppdspinfoam, Reisenbergerselfdual}, that guarantees the volume of a 4-simplex is the same for whatever pair of fluxes chosen to calculate it. To be more precise, for any pair of triangles $\tau$ and $\tau^\prime$ that belong to the same 4-simplex but lie in different tetrahedra, this constraint forces the quantity
\begin{equation}
\big|\epsilon_{\alpha\beta\mu\nu}\Sigma^{\alpha\beta}[\tau]\Sigma^{\mu\nu}[\tau^\prime]\big|
\end{equation}
to be independent of the pair of triangles under consideration. This constraint is automatically fulfilled once the 4-simplex is flat (and the Gauß law holds true). This has contributed to the believe that it could be ignored. We have  seen, in the last section, the wedges carry curvature. Therefore we cannot assume that the 4-simplices are flat, and should take the volume constraint more seriously.

We will now give a naive argument supporting the idea that the additional torsional constraint \eref{torscons} has something to do with the volume constraint just mentioned. Let us define the \qq{4-momentum of a tetrahedron}, i.e the four-vector
\begin{equation}
p^\alpha[\mathcal{T}]={}^3\mathrm{vol}[\mathcal{T}]n^\alpha[\mathcal{T}],
\end{equation}
in the frame of the center of the 4-simplex.
The volume ${}^4\mathrm{vol}[v]$ of a 4-simplex $v$ is a function of these momenta, it is in fact nothing but the wedge product of four of them, a short calculation gives
\begin{equation}
\big({}^4\mathrm{vol}[v]\big)^3=\pm\frac{3!}{4^3}\epsilon_{\alpha\beta\mu\nu}p^\alpha[\mathcal{T}_1]\,p^\beta[\mathcal{T}_2]\,p^\mu[\mathcal{T}_3]\,p^\nu[\mathcal{T}_4],\label{fourvol}
\end{equation}
with $\mathcal{T}_i$ for $i=1,\dots,5$ labelling the five tetrahedra bounding the 4-simplex. With the conservation law of the 4-momenta \eref{torscons} fulfilled, the 4-volume \eref{fourvol} would be obviously the same for whatever quadruple of tetrahedra is chosen (up to orientation). In this sense the additional torsional constraint \eref{torscons} has the same intention as the original volume constraint, and guarantees the volume of a 4-simplex is the same from whatever side we look at it.  We have thus found an additional constraint of torsional origin that should perhaps be included in the theory. 

The torsional equations \eref{trsnless}, \eref{gausstors} and \eref{torscons} should have an important geometrical interpretation provided by Minkowski's theorem \cite{Minkowskitheorem}. The Minkowski theorem holds in any dimension $N$, and states that given a number of co-vectors $v^1, \dots v^M$, $M>N$ that close to zero, there exists a unique $N$-dimensional convex polytope in $\R^N$, bounded by $N-1$-dimensional facets normal to $v^1, \dots v^M$, with their volume given by the magnitude of $v^1, \dots v^M$. The role of the Minkowski theorem for the three-dimensional geometry is well explored, \cite{twist, polyhdr, bianchisommer, pentahal}. 
The hope is, that the conservation law \eref{torscons} can provide the geometry of the spinfoam vertices, just as the Gauß law \eref{gausstors} uncovered the quantum-geometry at the nodes of the spin-network functions. This is a question that lies outside the scope of the present paper; a rigorous analysis of the role torsion plays for the geometry of the 4-simplex is missing, and we leave this task open for work to come.


\section{Quantum theory}\label{chapterIV}\noindent
The last section gave a Hamiltonian formulation of the dynamics of spinfoam gravity. On each edge we have introduced four twistorial fields $Z_{(I)}:[0,1]\ni t\mapsto (\bar\pi^{(I)}_{\bar A}(t),\omega^A_{(I)}(t))\in\T$, together with Lagrange multipliers $\lambda_{(I)}$ and $\ou{\Phi}{A}{B}(t)$. These variables must satisfy certain constraints: the simplicity constraints, the Gauß law, boundary conditions, and maybe also the additional torsional volume constraint introduced in section \ref{IIIH}.
The system has a finite number of degrees of freedom (four twistor fields, and a set of Lagrange multipliers), time-evolution is governed by a Hamiltonian (itself constrained to vanish), and the spinors are canonical (Darboux) coordinates covering all of phase-space. These key observations make quantisation rather straight forward.

The present section has two parts, in the first of which, section \ref{IVA}, we present the kinematical structure and solve the quantised constraints. Section \ref{IVB} concerns the dynamics of the system and defines the amplitudes for the case of a manifold without a boundary.
\subsection{Canonical quantisation and physical states}\label{IVA}
\noindent
The space of twistors $\T\ni Z=(\bar{\pi}_{\bar A},\omega^A)$ equipped with the Poisson brackets \eref{commtwist} can readily be quantised. Let us start with a brief review of what has already been shown in references \cite{komplexspinors} and \cite{twistintegrals}.  
Working with a \qq{position}-representation we are working on the Hilbert space $L^2(\C^2,d^4\omega)$. Wave-functions are non-analytic functions of the spinor and the integration measure is
\begin{equation}
d^4\omega=\frac{1}{16}\big(\di\omega_A\wedge\di\omega^A\wedge\CC\big).
\end{equation}
This Hilbert space is an auxiliary object, introduced just to have the proper arena to define the constraints. The \qq{position} operators $\hat\omega^A$ and $\hat{\bar{\omega}}^{\bar A}$ act by multiplication. The momentum, on the other hand, becomes a derivative:
\begin{equation}
\big(\hat{\pi}_A f\big)(\omega)=\frac{\hbar}{\I}\frac{\partial}{\partial\omega^A}f(\omega),\quad
\big(\hat{\bar{\pi}}_{\bar A} f\big)(\omega)=\frac{\hbar}{\I}\frac{\partial}{\partial{\bar\omega}^{\bar A}}f(\omega).
\end{equation}
On this Hilbert space the $SL(2,\C)$ action
\begin{equation}
\big(\mathcal{D}(g)f\big)(\omega^A)=f\big(\ou{(g^{-1})}{A}{B}\omega^B\big)
\end{equation}
is unitary, but reducible. Irreducible subspaces are spanned by distributions, that are homogenous in the spinors \cite{ruhl, unrepsl}. There are two quantum numbers $\rho\in\R$ and $k\in\frac{\Z}{2}$ labelling the irreducible subspaces $\mathcal{H}_{(\rho,k)}$, that parametrise the weight of homogeneity according to:
\begin{equation}
\forall z\in\C-\{0\}, f^{(\rho,k)}\in\mathcal{H}_{(\rho,k)}:f^{(\rho,k)}(z\omega^A)=z^{-k-1+\I\rho}{\bar z}^{+k-1+\I\rho}f^{(\rho,k)}(z\omega^A).
\end{equation}
This implies the important relations:
\begin{subequations}
\begin{align}
\omega^A\frac{\partial}{\partial\omega^A}f^{(\rho,k)}(\omega^A)&=(-k-1+\I\rho)f^{(\rho,k)}_{jm},\\
\bar\omega^{\bar A} \frac{\partial}{\partial\bar\omega^{\bar A}}f^{(\rho,k)}(\omega^A)&=(+k-1+\I\rho)f^{(\rho,k)}_{jm}.
\end{align}\label{eulerops}
\end{subequations}
We introduce the canonical quantisation of boosts and rotations:
\begin{equation}
\hat{\Pi}_i=-\ou{\tau}{AB}{i}\hat{\omega}_A\hat{\pi}_B,\quad \hat{L}_i=\hat{\Pi}_i+\HC,\quad\hat{K}_i=-\I\hat{\Pi}_i+\HC
\end{equation}
Notice no ordering ambiguity is appearing here, simply because the basis elements $\ou{\tau}{A}{Bi}$ are traceless $\ou{\tau}{A}{Ai}=0$.
We can work with a distributional basis on our auxiliary Hilbert space $L^2(\C^2,d^4\omega)$, that simultaneously diagonalises $\hat{L}_3$, $\hat{L}^2=\hat{L}^i\hat{L}_i$, together with the two Casimirs $\hat{L}^i\hat{K}_i$ and $\hat{L}^2-\hat{K}^2$ of the Lorentz group. The action of the Casimirs is in fact: 
\begin{equation}
\big(\hat{L}^2-\hat{K}^2+2\I \hat{L}_i\hat{K}^i\big)f^{(\rho,k)}_{jm}=-\hbar^2(\rho^2-k^2+1+2\I\rho k)f^{(\rho,k)}_{jm},
\end{equation}
which follows from \eref{eulerops} by expressing $\hat{\Pi}_i\hat{\Pi}^i$ in terms of $\hat{\omega}^A\hat{\pi}_{A}$.
Additional quantum numbers are spins $j=0,\tfrac{1}{2},1,\dots$ and the eigen-values $m=-j,\dots j$ of $\hat{L}_3$. The canonical basis reads explicitly:
\begin{equation}
f^{(\rho,k)}_{jm}(\omega^A)=\sqrt{\frac{2j+1}{\pi}}\|\omega\|^{2(\I\rho-1)}R^{(j)}\big(U^{-1}(\omega^A)\big){}^k{}_m,
\end{equation}
with
\begin{equation}
U(\omega^A)=\frac{1}{\|\omega\|^2}\begin{pmatrix}\omega^0 & -\bar{\omega}^{\bar 1}\\\omega^1&\phantom{-}\bar{\omega}^{\bar 0}\end{pmatrix}\in SU(2),
\end{equation}
and $R^{j}(U)^m{}_n=\bra{j,m}U\ket{j,n}$ being the Wigner matrix of the $j$-th irreducible $SU(2)$ representation.
The basis vectors obey the generalised orthogonality relations
\begin{equation}
\big\langle f^{(\rho,k)}_{jm},f^{(\rho^\prime,k^\prime)}_{j^\prime m^\prime}\big\rangle_{\C^2}=\int_{\C^2}d^4\omega\overline{f^{(\rho,k)}_{jm}(\omega^A)}f^{(\rho^\prime,k^\prime)}_{j^\prime m^\prime}(\omega^A)=\pi^2\delta(\rho-\rho^\prime)\delta_{kk^\prime}\delta_{jj^\prime}\delta_{mm^\prime}.
\end{equation}
We thus have a direct integral:
\begin{equation}
L^2(\C^2,d^4\omega)=\frac{1}{\pi}\int_\R^\oplus \di\rho\bigoplus_{k\in\Z}\mathcal{H}_{(\rho,k)}.
\end{equation}

We are now ready to discuss the quantisation of the constraints. For the Lorentz invariant part of the reality conditions we have to choose an ordering. For reasons that lie outside the scope of this paper we define it with the momentum and position operators in the following order:
\begin{equation}
\hat{D}=\frac{\hbar}{\beta+\I}\omega^A\frac{\partial}{\partial\omega^A}-\frac{\hbar}{\beta-\I}\Big(\bar{\omega}^{\bar A}\frac{\partial}{\partial{\bar\omega}^{\bar A}}+2\Big)
\equiv\frac{\hbar}{\beta+\I}\omega^A\frac{\partial}{\partial\omega^A}+\HC
\end{equation}
This operator is diagonal on the homogenous functions. With the action of the Euler operators \eref{eulerops} we get immediately:
\begin{equation}
\hat{D}f^{(\rho,k)}_{jm}=\frac{2\hbar}{\beta^2+1}\big(\rho-\beta(k+1)\big)f^{(\rho,k)}_{jm}.
\end{equation}
The solution space of this constraint is non-normalisable in $L^2(\C^2,d^4\omega)$, simply because we cannot integrate the homogenous functions along the rays $\omega^A(z)=z\omega^A(0)$ (with $z\in\C$). We can, however, introduce a surface integral removing this divergence. Take the Hamiltonian vector field $\mathfrak{X}_D$, defined for any scalar function $f$ as $\mathfrak{X}_D[f]=\{D,f\}$, and consider the interior product $\iota_{\mathfrak{X}_D}d^4\omega$. This defines a non-degenerate 3-form on the space of orbits \eref{dtrafos} generated by $D$, that we are using to define an inner product:
\begin{equation}
\big\langle f,f^\prime\big\rangle_{\C^2/D}=\frac{\beta^2+1}{2\pi}\int_{\C^2/D}(\iota_{\mathfrak{X}_D}d^4\omega)\,\overline{f}f^\prime.\label{Dprod}
\end{equation}
To evaluate this integral we need to choose a gauge section, that embeds the three-dimensional surface $\C^2/D$ into $\C^2$. The inner product is independent of this choice only if the Lie derivative of the integrand vanishes:
\begin{equation}
\mathcal{L}_{\mathfrak{X}_D}\big(d^4\omega\,\overline{f}f^\prime\big)=0
\end{equation}
If both $f$ and $f^\prime$ are in the kernel of $\hat{D}$ this condition holds true, and we arrive at a well defined inner product. In that case, we can deform the integration domain in the direction of $\mathfrak{X}_D$, that is along the orbits generated by $D$, without changing the integral. For our basis vectors we get in fact:
\begin{equation}
\big\langle f^{(\beta(k+1),k)}_{jm},f^{(\beta(k^\prime+1),k^\prime)}_{j^\prime m^\prime}\big\rangle_{\C^2/D}=\frac{\beta^2+1}{2\pi}\int_{\C^2/D}(\iota_{\mathfrak{X}_D}d^4\omega)\,\overline{f^{(\beta( k+1),k)}_{jm}}f^{(\beta(k^\prime+1),k^\prime)}_{j^\prime m^\prime}=\delta_{kk^\prime}\delta_{jj^\prime}\delta_{mm^\prime},\label{orthrel}
\end{equation}
implicitly showing the integral \eref{Dprod} defines a nondegenerate inner product on the kernel of $\hat{D}$.
We are now left with the remaining $F_n=0$ constraint. Knowing the classical constraint generates an additional $\su(2)$ algebra (remember equation \eref{hiddensu2}) it is not hard to see that the operators
\begin{equation}
\hat{F}_n=n^{A\bar A}\hat{\bar\pi}_{\bar A}\hat{\omega}^A,\quad
\hat{\bar F}_n=\hat{F}^\dagger_n=n^{A\bar A}\hat{\pi}_{A}\hat{\bar\omega}^{\bar A},
\end{equation}
act as creation and annihilation operators for the quantum number $k$, more explicitly:
\begin{subalign}
\hat{F}_{n_o}&f^{(\rho,k)}_{jm}=-\frac{\hbar}{\sqrt{2}}\sqrt{(j-k)(j+k+1)}f^{(\rho,k+1)}_{jm},\\
\hat{F}^\dagger_{n_o} &f^{(\rho,k)}_{jm}=-\frac{\hbar}{\sqrt{2}}\sqrt{(j+k)(j-k+1)}f^{(\rho,k-1)}_{jm}.
\end{subalign}
Here we have chosen time-gauge where the normal $n_o^{A\bar A}$ assumes the form of \eref{tgauge}. Unless $j=0=k$ we cannot find states in the kernel of both $\hat{F}_n$ and its Hermitian conjugate, which reflects $F_n$ is of second class. We proceed with Gupta and Bleuler \cite{gupta, bleuler} and impose only one half of the constraint. The kernel of $\hat{F}_n$ is spanned by states $k=j$:
\begin{equation}
\hat{F}_{n_o}f^{(\rho,j)}_{jm}=0,\label{sol1}
\end{equation}
while $k=-j$ labels the states in the kernel of its Hermitian conjugate:
\begin{equation}
\hat{F}_{n_o}^\dagger f^{(\rho,-j)}_{jm}=0.\label{sol2}
\end{equation}
We can restrict ourselves to only one of these two possibilities. This is motivated as follows. The quantum number $k$ is an eigenvalue of the operator $\hat{\omega}^A\hat{\pi}_A+\hat{\pi}_A\hat{\omega}^A-\HC$, we have
\begin{equation}
\big[\hat{\omega}^A\hat{\pi}_A+\hat{\pi}_A\hat{\omega}^A-\HC\big]f^{(\rho,k)}_{jm}=4\I\hbar k f^{(\rho,k)}_{jm}.
\end{equation}
This operator represents the following classical quantity on the solution space of the simplicity constraints:
\begin{equation}
2\omega^A\pi_A-\CC=4\I J.
\end{equation}
Just below equation \eref{Jinterpret} we argued that we can always assume $J>0$ thereby removing the discrete symmetry exchanging $\omega$ and $\pi$. If we agree on the constraint $J>0$ also in quantum theory we can discard the solution \eref{sol2} and just work with \eref{sol1}. The solution space of the simplicity constraints $\mathcal{H}_{\mathrm{simpl}}$ is then restricted to the kernel of the operators $\hat{D}$ and $\hat{F}_{n_o}$, it is spanned by the basis vectors $f^{(\beta(j+1),j)}_{jm}$, and we so have
\begin{equation}
\mathcal{H}_{\mathrm{simpl}}=\overline{\mathrm{span}}\big\{f^{(\beta (j+1),j)}_{jm}: j=0,\tfrac{1}{2},\dots;\,m=-j,\dots ,j\big\}.
\end{equation} 
The Gauß constraint is the last remaining constraint to impose. As mentioned in above we only need to solve the rotational part of the Gauß law, which appears as a first-class constraint in our set of constraints \eref{setcons}. 
The canonical quantisation of the classical constraint becomes the operator 
\begin{equation}
\hat{G}_i^{\mathrm{rot}}=\sum_{i=1}^4\hat{L}_i^{(I)}.
\end{equation}
on $\bigotimes^4L^2(\C^2,d^4\omega)$, with e.g.:
\begin{equation}
\hat{L}_i^{(1)}=\hat{L}_i\otimes\mathds{1}\otimes\mathds{1}\otimes\mathds{1}.
\end{equation}
The constraint generates rotations leaving invariant the 4-normal $n_o$, e.g. for $f\in\bigotimes^4L^2(\C^2,d^4\omega)$:
\begin{equation}
\big(\exp(-\tfrac{\I}{\hbar}\hat{G}_i^{\mathrm{rot}}\varphi^i)f\big)(\omega_{(1)},\dots,\omega_{(1)})=
f(U\omega_{(1)},\dots,U\omega_{(4)}),\quad\text{with:}\; U=\exp(\varphi^i\tau_i)\in SU(2).
\end{equation}
The kernel of all constraints, the Hilbert space $\mathcal{H}_{\mathrm{phys}}$, is thus given by the $SU(2)$ invariant part of $\bigotimes^4\mathcal{H}_{\mathrm{simpl}}$, that is:
\begin{equation}
\mathcal{H}_{\mathrm{phys}}=\Big(\bigotimes^4\mathcal{H}_{\mathrm{simpl}}\Big)/{SU(2)}.
\end{equation}
The general state in this Hilbert space looks like this:
\begin{equation}
\begin{split}
\Psi(\omega_{(1)},\omega_{(2)},\omega_{(3)},\omega_{(4)})=\sum_{m_1=-j_1}^{j_1}&\dots\sum_{j_,\dots,j_4}\sum_{m_4=-j_4}^{j_4} I^{m_1\dots m_4}(j_1,\dots,j_4)
f^{(\beta(j_1+1),j_1)}_{j_1m_1}(\omega_{(1)})\\
f^{(\beta(j_2+1),j_2)}_{j_2m_2}(\omega_{(2)})
&f^{(\beta(j_3+1),j_3)}_{j_3m_3}(\omega_{(3)})
f^{(\beta(j_4+1),j_4)}_{j_4m_4}(\omega_{(4)}),
\end{split}\label{physstate}
\end{equation}
with $I^{m_1\dots m_4}(j_1,\dots,j_4)$ being an intertwiner, which is an element of the spin zero component of the tensor product of $SU(2)$ representations of spins $j_1,\dots, j_4$. The defining property of an intertwiner is that it be $SU(2)$ invariant:
\begin{equation}
\forall U\in SU(2):\sum_{n_1=-j_1}^{j_1}\dots\sum_{n_4=-j_4}^{j_4}I^{n_1\dots n_4}(j_1,\dots,j_4)\ou{R^{(j_1)}(U)}{m_1}{n_1}\dots \ou{R^{(j_4)}(U)}{m_4}{n_4}=I^{m_1\dots m_4}.
\end{equation}

Before we go on to the next section, let us comment on how to relax time-gauge.  For any two normals $n$ and $n_o$ on the upper hyperboloid $\eta_{\alpha\beta}n^\alpha n^\beta=-1,\; n^0>0$ there is a proper orthochronous Lorentz transformation that sends one to the other. Let $g_n$ be the corresponding $SL(2,\C)$ element such that:
\begin{equation}
n^{A\bar A}=\uo{g_n}{B}{A}\uo{\bar{g}_n}{\bar B}{\bar A}n^{B\bar B}_o
\end{equation}
We then also have:
\begin{equation}
\hat{F}_n=\mathcal{D}(g_n)^{-1}\hat{F}_{n_o}\mathcal{D}(g_n).
\end{equation}
Any vector in the kernel of $\hat{F}_n$ can be constructed from its preimage in $\mathcal{H}_{\mathrm{simpl}}$. The vectors
\begin{equation}
f^{(\beta (j+1),j)}_{jm(n)}:=\mathcal{D}(g_n)^{-1}f^{(\beta (j+1),j)}_{jm}
\end{equation}
are in fact an orthonormal basis in the kernel of $\hat{F}_n$. More importantly, the constraints $\hat{F}_n$ and $\hat{F}_n^\dagger$ weakly vanish. All matrix elements of $\hat{F}_n$ with respect to states $f^{(\beta(j+1),j)}_{jm(n)}$ and $f^{(\beta(j^\prime+1),j)}_{j^\prime m^\prime(n^\prime)}$ equate to zero:
\begin{equation}
\big\langle f^{(\beta (j+1),j)}_{jm(n)},\hat{F}_{n}f^{(\beta (j^\prime+1),j^\prime)}_{j^\prime m^\prime(n^\prime)}\big\rangle_{\C^2}=0=\big\langle f^{(\beta (j+1),j)}_{jm(n)},\hat{F}_{n^\prime}f^{(\beta (j^\prime+1),j^\prime)}_{j^\prime m^\prime(n^\prime)}\big\rangle_{\C^2}.\label{weakcons}
\end{equation}
As before, we are now left to impose the rotational part of the Gauß constraint. But we have left time gauge, and the $SU(2)$-Gauß law becomes boosted to:
\begin{equation}
\hat{G}_{i(n)}^{\mathrm{rot}}=\mathcal{D}(g_n)^{-1}\hat{G}_{i}^{\mathrm{rot}}\mathcal{D}(g_n).
\end{equation}
The general solution $\Psi_{(n)}$ of all constraints can thus easily be found from \eref{physstate} by just performing a unitary transformation:
\begin{equation}
\Psi_{(n)}(\omega_{(1)},\dots,\omega_{(4)})=\big(\mathcal{D}(g_n)^{-1}\Psi\big)(\omega_{(1)},\dots,\omega_{(4)})=
\Psi(g_n\omega_{(1)},\dots,g_n\omega_{(4)})
\end{equation}
These states are nothing but the spinorial equivalent of Levine's projected spin-network states \cite{liftng, projspinnet}.
\subsection{Local Schrödinger equation and spinfoam amplitude}\label{IVB}\noindent
To begin with, consider only the evolution of the quantum states along a single edge. As in the classical part of the paper we can align the space-time normal in the middle of the edge to the canonical choice, that is we go to the time gauge \eref{tgaugeedge} at $t=t_o=\tfrac{1}{2}$.

Classically, the Hamilton function governs the time evolution along an edge. An observable $O_t:\T\rightarrow\R$ on the phase space of a single triangle evolves according to
\begin{equation}
\frac{\di}{\di t}O_t=\Big\{\big(\Phi^{AB}(t)\pi_A\omega_B+\CC\big)+\lambda(t)D,O_t\Big\}.
\end{equation}
With $\Phi^{AB}(t)$ again being the selfdual connection contracted with the tangent vector of the edge, as defined in \eref{lagmult}. When going to the quantum theory the Hamiltonian function becomes an operator defining the Schrödinger equation:
\begin{equation}
\I\hbar\frac{\di}{\di t}\psi_t=\big(\Phi^{AB}(t)\hat{\pi}_A\hat{\omega}_B+\HC\big)\psi_t+\lambda(t)\hat{D}\psi_t.\label{schroed}
\end{equation}
This is an important intermediate result. The Hamiltonian on the right hand side, agrees with what Bianchi has reported in his thermodynamical considerations of spinfoam gravity \cite{Bianchientropy}. If we restrict $\Phi$ to be a boost in the direction orthogonal to the triangle, we end up with the \qq{boost-Hamiltonian} \cite{bwsurf}, that becomes the energy of a locally accelerated observer \cite{Perezlocal} once we are in the semi-classical regime.
 
At $t=t_o$ we are in the time gauge, physical states are annihilated by  $\hat{F}_{n_o}$, and lie in the kernel of $\hat{D}$, such that our initial condition becomes
\begin{equation}
\psi_{t=t_o}=\sum_{j=0}^\infty\sum_{m=-j}^jc^{jm}f^{(\beta(j+1),j)}_{jm}.
\end{equation}
The last part of $\eref{schroed}$ vanishes on the physical Hilbert space, implying the Hamiltonian acts as an infinitesimal Lorentz transformation. We have:
\begin{equation}
\begin{split}
\big(\Phi^{AB}\hat{\pi}_A\hat{\omega}_B+\HC\big)f(\omega^A)&=
\I\hbar\big(\ou{\Phi}{A}{B}\omega^B\frac{\partial}{\partial\omega^A}+
\ou{\bar\Phi}{\bar A}{\bar B}\bar\omega^{\bar B}\frac{\partial}{\partial\bar\omega^{\bar A}}\big)f(\omega^A)=\\
&=\I\hbar\frac{\di}{\di\varepsilon}\Big|_{\varepsilon=0}f\big(\ou{\exp(\varepsilon \Phi)}{A}{B}\omega^B\big).
\end{split}
\end{equation}
We can thus trivially integrate the Schrödinger equation to find:
\begin{equation}
\psi_t(\omega^A)=\sum_{j=0}^\infty\sum_{m=-j}^jc^{jm}f^{(\beta(j+1),j)}_{jm}\big(\ou{U}{A}{B}(t,t_o)\omega^B\big)=\big(\mathcal{D}(U(t_o,t))\psi_{t_o}\big)(\omega^A).
\end{equation}
Where we have inserted the $SL(2,\C)$ holonomy along the edge, introduced in \eref{edgehol}. With the normal parallel along the edge, hence transported by the holonomy as in \eref{normtrans}, equation \eref{weakcons} implies the constraint $F_n=0$ holds weakly for all times, i.e.:
\begin{equation}
\forall f\in \mathrm{kern}(\hat{D}), t\in[0,N):\big\langle f,\hat{F}_{n(t)}\psi_t\big\rangle_{\C^2}=0=\big\langle f,\hat{F}_{n(t)}^\dagger\psi_t\big\rangle_{\C^2}.
\end{equation}
Consider now the process where the spinor is \qq{scattered} from $t_o=\tfrac{1}{2}$ into $t_1=\tfrac{3}{2}$ passing through a vertex. Since we are both at $t_o$ and $t_1$ in the canonical gauge \eref{tgaugemany}, we can take as initial and final states
\begin{equation}
\psi^{\mathrm{final}}_{t_1}=f^{(\beta(j^\prime+1),j^\prime)}_{j^\prime m^\prime},\quad \psi^{\mathrm{initial}}_{t_o}=f^{(\beta(j+1),j)}_{jm}.
\end{equation}
The corresponding transition amplitude is: 
\begin{equation}
A(\psi^{\mathrm{initial}}_{t_o}\rightarrow\psi^{\mathrm{final}}_{t_1})=\Big\langle f^{(\beta (j^\prime+1),j)}_{j^\prime m^\prime},\mathcal{D}(U(t_o,t_1))f^{(\beta(j+1),j)}_{jm}\Big\rangle_{\C^2/D},\label{transampl}
\end{equation}
which vanishes unless $j=j^\prime$ due to \eref{orthrel}. With $j$ being the quantisation of $J$, as defined in \eref{Jinterpret}, we see, also in quantum theory, the area of the triangle is preserved when going around the spinfoam face. This is the quantum theoretical version of the area-matching-constraint $\dot{E}=0$ introduced in \eref{conslaw}. 

We are now going to close the edges to form a loop, obtaining the amplitude\footnote{In quantum mechanics, the analogue of what we are calculating here, is the \qq{partition} function $Z(\I t)=\mathrm{Tr}(\E^{-\I t\hat{H}})$. } for a spinfoam face $f$. The boundary of the spinfoam face passes through vertices $v_1,\dots,v_{N}$ lying between edges $\{e_i\}_{i=1,\dots N}$ that go from the vertex $v_{i-1}$ towards the $i$-th. By going around the spinfoam face we will see $N$ processes of the form of \eref{transampl} happening. We write the elementary amplitude for the scattering process \eref{transampl} at the $i$-th vertex in the condensed form of
\begin{equation}
\bra{jm_{i+1}}g_{e_i,e_{i+1}}\ket{jm_i}=\Big\langle f^{(\beta (j+1),j)}_{jm_{i+1}},\mathcal{D}(U(\tfrac{2i-1}{2},\tfrac{2i+1}{2}))f^{(\beta(j+1),j)}_{jm_i}\Big\rangle_{\C^2/D},
\end{equation}
where we used the abbreviations
\begin{equation}
g_{e_i,e_{i+1}}=g_{e_{i+1}}^{\mathrm{source}}(g_{e_i}^{\mathrm{target}})^{-1},\quad g_{e_i}^{\mathrm{target}}=U(i,\tfrac{2 i-1}{2}),\quad g_{e_i}^{\mathrm{source}}=U(i-1,\tfrac{2 i-1}{2}).
\end{equation}
Here $i=N+1$ has everywhere to be identified with $i=1$. We obtain the amplitude $Z_f$, i.e. the \qq{partition} function for a spinfoam face $f$, by summing the product of the amplitudes for each individual process \eref{transampl} over the orthonormal basis at the edges, that is we have to trace over spins $j$ and $m_{i=1\dots N}=-j,\dots, j$. The resulting quantity depends parametrically on the edge holonomies $g_{e_i,e_{i+1}}$ as follows:
\begin{equation}
Z_f(\underline{g})\equiv Z_f(g_{e_0,e_1},\dots,g_{e_{N-1},e_0})=\sum_{j=0}^\infty\sum_{m_1=-j}^{j}\dots\sum_{m_N=-j}^{j}\prod_{i=1}^{N}\bra{jm_{i+1}}g_{e_i,e_{i+1}}\ket{jm_i}.
\end{equation}
This amplitude is in exact agreement with the EPRL-model. To obtain the spinfoam amplitude for the whole discretised space-time manifold we take the product of all $Z_f$ over all individual spinfoam faces $f$ and integrate over the free gauge parameters left. These are the edge holonomies $g_e^{\mathrm{source}}$ and $g_e^{\mathrm{target}}$. We want to ensure local Lorentz invariance \cite{lortzcov}, ans thus take the integration measure to be the Haar measure $dg$ of $SL(2,\C)$. This measure is unique up to an overall constant. The resulting quantity matches the EPRL model. 

What we would now like to propose is a modification of the EPRL-model that takes into account the additional torsional constraint discussed in \eref{torscons}. The most naive way to impose the conservation law \eref{torscons} would just involve a delta function of the constraint inserted at each vertex of the spinfoam amplitude:
\begin{equation}
Z=\prod_{e:\text{edges}}\int_{SL(2,\C)}\!\!\!\!dg_e^{\text{source}}\int_{SL(2,\C)}\!\!\!\!dg_e^{\text{target}}\sum_{\eta_e\in\{-1,1\}}\prod_{v:\text{vertices}}\delta_{\R^4}\Big(\sum_{\mathcal{T}\in v}\varepsilon[\mathcal{T},\eta_{e(\mathcal{T})}]n^\alpha[\mathcal{T}]\,{}^3\widehat{\mathrm{vol}}[\mathcal{T}]\Big)\prod_{f:\text{faces}}Z_f(\underline{g})\label{ampltde}
\end{equation}
Here $n^\alpha[\mathcal{T}]$ is the normal of the tetrahedron $\mathcal{T}$ parallel transported into the center of the 4-simplex, ${}^3\widehat{\mathrm{vol}}[\mathcal{T}]$ denotes the quantisation of its volume \eref{3vol}, $e(\mathcal{T})$ is the edge dual to $\mathcal{T}$, while $\varepsilon[\mathcal{T},\eta_{e(\mathcal{T})}]$ gives the orientation of the tetrahedron relative to the vertex it is seen from. This sign tells us whether the outwardly pointing normal of the boundary of the 4-simplex is future (i.e. $\varepsilon=+1$) or past (i.e. $\varepsilon=-1$) oriented. 

We will now give the missing definitions for the orientation $\varepsilon[\mathcal{T},\eta]$ and the time normals $n^\alpha[\mathcal{T}]$. Let $\mathcal{T}^{\mathrm{target}}$ and $\mathcal{T}^{\mathrm{source}}$ be the same tetrahedron seen from vertices $v^{\mathrm{target}}$ and $v^{\mathrm{source}}$, and the intermediate edge $e(\mathcal{T}^{\mathrm{source}})=e(\mathcal{T}^{\mathrm{target}})$ be oriented from $v^{\mathrm{source}}$ towards $v^{\mathrm{target}}$. We define the orientation by setting:
\begin{equation}
\varepsilon[\mathcal{T}^{\mathrm{source}},\pm 1]=\pm 1,\quad \varepsilon[\mathcal{T}^{\mathrm{target}},\pm 1]=\mp 1.\label{epsdef}
\end{equation} 
The time normals are given by equation \eref{normtrans} implying:
\begin{subalign}
\frac{\I}{\sqrt{2}}\delta^{A\bar A}&=\uo{\big(g^{\mathrm{target}}_e\big)}{B}{A}\uo{\big(\bar{g}^{\mathrm{target}}_e\big)}{\bar B}{\bar A}n^{B\bar B}[\mathcal{T}^{\mathrm{target}}],\\
\frac{\I}{\sqrt{2}}\delta^{A\bar A}&=\uo{\big(g^{\mathrm{source}}_e\big)}{B}{A}\uo{\big(\bar{g}^{\mathrm{source}}_e\big)}{\bar B}{\bar A}n^{B\bar B}[\mathcal{T}^{\mathrm{source}}].
\end{subalign}

It definitely exceeds the scope of this paper to give a detailed analysis of the amplitude \eref{ampltde} proposed. But let us make two immediate observations and comments. The first concerns causality. The function $\varepsilon$ defined in \eref{epsdef} assigns to any tetrahedron a local time orientation, and tells us whether the outwardly pointing 4-normal of a tetrahedron bounding a 4-simplex is future or past oriented---that is, so to say, whether the tetrahedron \qq{enters} or \qq{leaves} the 4-simplex.   
This would distinguish 4-simplices corresponding to 3-1 (1-3) moves from those representing 4-1 (1-4) moves, which could eventually introduce a notion of causality. 
The second remark is about the asymptotic analysis of Barret et. al. \cite{asymptcs, Hanasymptotics}. They show the EPRL vertex amplitude reproduces the Regge action if the quantum state in the boundary Hilbert space is peaked on area-angle variables representing a flat 4-simplex in Minkowski space. I find it quite plausible that the additional delta function restricts the quantum geometries appearing in the amplitude \eref{ampltde} to exactly those that fulfil this criterion.

\section{Conlusion}\noindent
To canonically quantise gravity it is often thought that one first needs to start from a 3+1 spilt, study the ADM (Arnowitt--Deser--Miser) formulation in the \qq{right} variables, identify the canonical structure and perform a Schrödinger quantisation. This paper questioned this idea. The ADM formulation is very well adapted to a continuous space-time, but in spinfoam gravity we are working with a discretisation of the manifold, hence lacking that assumption. Instead we have simplices glued together and should find a Hamiltonian formulation better adapted to the problem. This is what the paper achieves in the classical part. We found a Hamiltonian for the discretised theory, generating the time evolution along the edges of a spinfoam. The time variable introduced is nothing but a coordinate, and does not measure duration as given by a clock.

We started from a topological theory, and took the spinorial framework of loop quantum gravity to parametrise the action. This we did for technical reasons only, spinors do not add anything physically new to the theory. The key idea was to perform a limiting process that partially returns to the continuum. We split every wedge into smaller and smaller parts, until we obtained a continuum action on an edge. Next, we have added the simplicity constraints to the action. The equations of motion allowed for a Hamiltonian formulation, such that we could go through the Dirac analysis of the constraint algebra. This revealed all constraints are preserved in time (i.e. along the boundary of the spinfoam face) provided the Lagrange multiplier in front of the second-class constraint $F_n=0$ vanishes. 

Then we studied the equations of motion for the spinors. We found they can easily be integrated, the only trouble being the periodic boundary conditions, that imply a constraint on the holonomy along the loop bounding the spinfoam face, i.e. equation \eref{reggehol}. This parallel transport is neither a pure boost, as in Regge calculus, nor a rotation, but a combination of both, with the Barbero--Immirzi parameter measuring the relative strength. Nevertheless, there are key similarities with Regge calculus. If parallel transported along the bounding loop, the flux through the triangle dual to the spinfoam face is mapped into itself, while the curvature \eref{curvtens} is a function of the deficit angles between adjacent tetrahedra \eref{lambdainterprett}. 
 
The classical part concluded with a reflection on the role of torsion in a discrete theory of gravity. We saw, torsion not only implies the Gauß law for each tetrahedron, but also an additional constraint \eref{torscons} on each vertex of the simplicial decomposition.  This constraint demands, that on every 4-simplex the outwardly pointing normals of the bounding tetrahedra weighted by their volumes sum up to zero. We stressed similarities of this additional torsional constraint with the volume constraint. The volume constraint together with the Gauß and the linear simplicity constraints guarantee the existence of a tetrad around the 4-simplex. Assuming the 4-simplex were flat, this constraint could be ignored. We questioned this claim by stressing that the model carries curvature and the 4-simplices generically cannot taken to be flat.

We introduced the additional torsional constraint essentially by hand. If it turns out the theory cannot work without this constraint, it must arise more naturally, e.g. from an additional term in the action. Our discussion on torsion therefore clearly calls for a more systematic analysis, that should be based on what has already been sketched in section \eref{IIIH}. This is left for a project yet to come. 

In summary, the classical part introduced a canonical formulation of spinfoam gravity adapted to a simplicial discretisation of space-time. This framework should be of general interest, as it provides a solid foundation where different models could fruitfully be compared.

The last section was about the quantum theory. With the Hamiltonian formulation of the spinfoam dynamics at hand, canonical quantisation was straight-forward. We used an auxiliary Hilbert space to define the operators. Physical states are in the kernel of the first-class constraints. The second-class constraints act as ladder operators. One of them ($\hat{F}_n$) annihilates physical states, while the other one ($\hat{F}^\dagger_n$) maps them to their orthogonal complement, i.e. into the spurious part of the auxiliary Hilbert space. This is exactly what happens in the Gupta--Bleuler formalism. 

In quantum theory, dynamics is determined by the Schrödinger equation. We quantised the classical Hamiltonian and solved the Schrödinger equation that gives the evolution of the quantum states along the boundary of a spinfoam face. This boundary evolution matched the Schrödinger equation introduced by Bianchi in the thermodynamical analysis of spinfoam gravity \cite{Bianchientropy}. 
Gluing the individual transition amplitudes together, we got the amplitude for a spinfoam face, which was in exact agreement with the EPRL model. The only departure from the EPRL model, concerned the additional torsional constraint we added to the amplitude. We argued for two advantages this could provide. First, it could introduce a notion of causality. This may happen, since the torsional constraint distinguishes the future-pointing from the past-pointing normals of the boundary of the 4-simplex. The second possible advantage concerns the asymptotic behaviour of the amplitude. Classically, the additional torsional constraint should restrict the area-angle variables on the boundary to those that close to form 4-simplex. These are called Regge-like boundary data in the terminology of \cite{asymptcs, Hanasymptotics}. If the torsional constraint can achieve the same also in the quantum theory, it would be important for us, because for those boundary data the spinfoam amplitude is already known to reproduce the Regge action, which, in turn, has the correct continuum limit.
\newline
\paragraph*{Acknowledgements} I thank Eugenio Bianchi, Hal Haggard, Muxin Han, Carlo Rovelli, Simone Speziale and Edward Wilson--Ewing for enlightening discussions on the draft of this paper, and Hal Haggard also for a careful reading of the draft.


\begin{flushleft}

\end{flushleft}

\end{document}